# Introduction to Beam Instrumentation and Diagnostics

*M. Gasior, R. Jones, T. Lefevre, H. Schmickler*
CERN, Geneva, Switzerland

*K. Wittenburg*
DESY, Hamburg, Germany

**Abstract**
These lectures aim to describe instruments and methods used for measuring beam parameters in particle accelerators. Emphasis will be given to new detection and analysis techniques in each field of accelerator instrumentation. A clear distinction is made between 'instrumentation', the design and construction of the instruments themselves, and 'diagnostics', the use of the data from these instruments for running and improving the performance of the accelerator.

## 1    Introduction

Beam instrumentation and diagnostics combine the disciplines of accelerator physics with mechanical, electronic, and software engineering, making it an extremely interesting field in which to work. The aim of the beam instrumentation physicist or engineer is to design, build, maintain, and improve the diagnostic equipment for the observation of particle beams with the precision required to tune, operate, and improve the accelerators and their associated transfer lines.

This introduction is intended to give an overview of the instrumentation in use in modern accelerators. The choice available today is so vast that inevitably it will not be possible to cover them all. Many of the standard instruments have been covered in previous CAS schools (see for example [1], which also contains a comprehensive list of references) and will therefore be touched upon only briefly. The following subjects will be discussed: beam position measurement; beam current and intensity measurement; diagnostics of transverse beam motion (tune, chromaticity, and coupling); emittance measurement; beam loss monitoring; longitudinal profile measurement; and some examples of beam diagnostics

## 2    Beam position measurement

The beam position monitor (BPM) can be found in every accelerator. Its role is to provide information on the position of the beam in the vacuum chamber at the monitor location. For linacs and transfer lines the BPMs are used to measure and correct beam trajectories, while for synchrotrons such monitors are distributed around the ring and used to calculate the closed orbit. In circular machines, their location is usually chosen close to the main quadrupole magnets where the β-functions are largest and so any orbit distortion is at a maximum. For 90° lattices a typical layout involves placing horizontal monitors near the focusing quadrupoles (where the horizontal β-function is large) and the vertical monitors near the defocusing quadrupoles (where the vertical β-function is large). Apart from closed orbit measurements, the BPMs are also used for trajectory measurements (the first turn trajectory is particularly important for closing the orbit on itself) and for accelerator physics experiments, where turn-by-turn data, and even bunch-to-bunch data is often required.

In the early days a BPM monitoring system simply consisted of an oscilloscope linked directly to the pick-up signals. Since then, enormous advances in the acquisition and processing electronics have been made, turning BPMs into very complex systems. Modern BPMs are capable of digitizing individual bunches separated by a few nanoseconds, with a spatial resolution of a micrometre or less, while the resulting orbit or trajectory collected from several hundred pick-ups can be displayed in a fraction of a second.

## 2.1 Pick-ups

The measurement of beam position relies on processing the information from pick-up electrodes located in the beam pipe. Five pick-up families are commonly employed:

i) electrostatic—including so-called 'button' and 'shoe-box' pick-ups;

ii) electromagnetic—stripline couplers;

iii) resonant cavity—especially suited for high-frequency linacs;

iv) resistive;

v) magnetic.

An excellent in-depth analysis of most of these pick-ups is presented in Ref. [2]. Here we will briefly describe the three most commonly used, namely the electrostatic, electromagnetic, and cavity pick-ups.

### 2.1.1 *Electrostatic (capacitive)*

The electrostatic or capacitive pick-up is the most widely used in circular accelerators. It consists of metallic electrodes situated on opposite sides of the vacuum chamber at the location where the beam is to be measured. As the beam passes through, electric charges are induced on the electrodes, with more induced on the side that is closer to the beam than the one furthest from the beam. By measuring the difference in the charge induced, the position can be calculated.

Let us analyse the properties of button pick-ups (see Fig. 1) since they are the most popular due to their low cost and ease of construction.

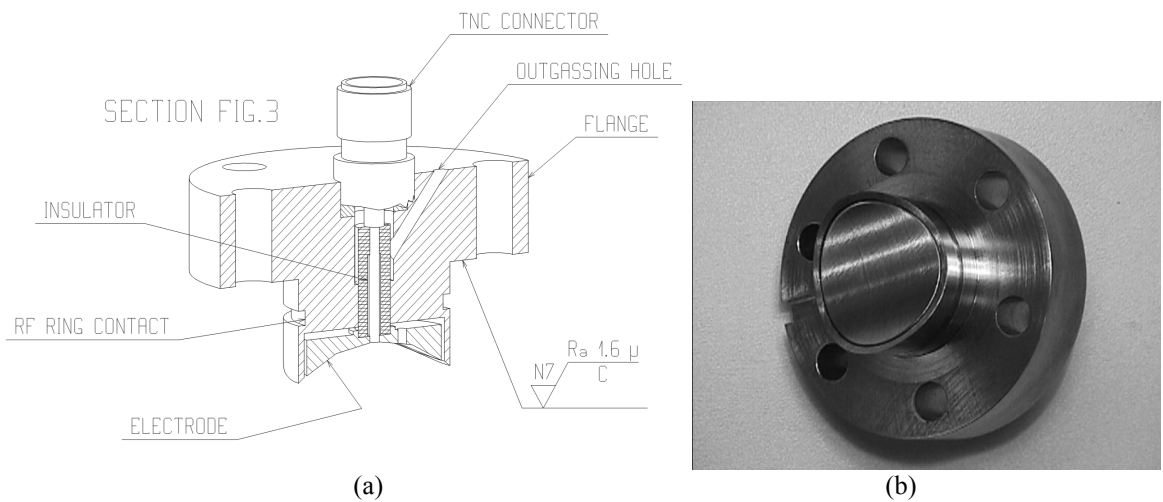

(a) (b)

**Fig. 1:** (a) Cross-section through and (b) photo of an LHC button electrode

The image current associated with the beam will induce a charge on the button that is proportional to the beam intensity and inversely proportional to the position of the beam from the electrode. A schematic representation is given in Fig. 2.

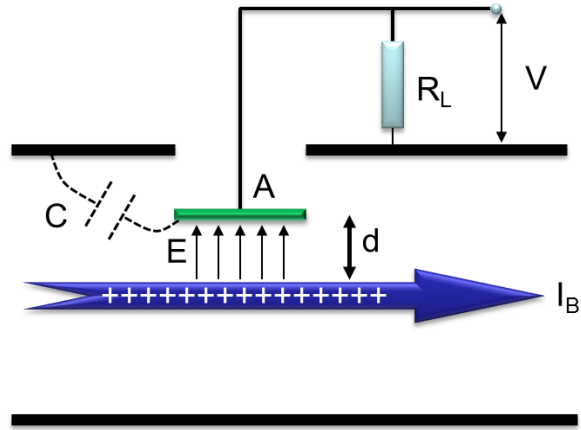

**Fig. 2:** Schematic of a capacitively coupled electrode

The figure of merit for any electrode is its transfer impedance (the ratio of the pick-up output voltage ($V$) to the beam current ($I_B$)). For a capacitive pick-up the signal is proportional to the rate of change of beam current at low frequencies, while for high frequencies the capacitance 'integrates' the signal and the transfer impedance tends to its maximum. For the case of a button electrode of area $A$ and capacitance $C$ situated at a distance $d$ from the beam, the maximum transfer impedance (i.e. the value it tends to at high frequency) can be approximated by

$$Z_{T\infty} = \frac{A}{2\pi d (\beta c) C} \tag{1}$$

where $\beta$ is the relativistic $\beta$ and $c$ the speed of light.

Impedance transformation can be used to improve the low-frequency response at the expense of that at high frequency. Figure 3(a) shows the frequency response of an 8 pF button electrode for the case of matched 50 Ω impedance (1:1) and after two different impedance transformations. The time response of the button for different bunch lengths can be seen in Fig. 3(b).

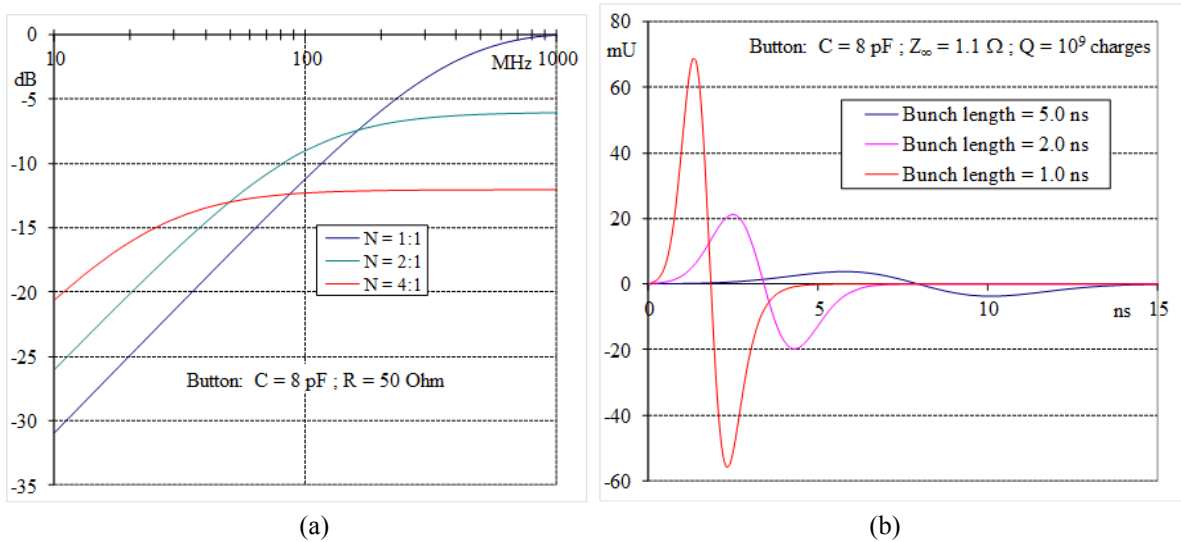

**Fig. 3:** (a) Frequency response and (b) time response of a button electrode

When designing such pick-ups care must be taken to limit the impedance variations when the transmission line used for signal extraction passes from vacuum to a feedthrough or cable dielectric (such as ceramic, glass or air). Any such mismatch will produce unwanted reflections, often at high frequency, which could perturb the processing electronics. For this reason most processing chains introduce a low-pass filter on the button output. Special care must also be taken to pair the electrodes on opposite sides of the chamber to minimize offsets in the position reading. This pairing can be made less sensitive to capacitance variations if the high-frequency cut-off for the processing electronics sits on the linear part of the button response, with the disadvantage that the overall signal amplitude is reduced.

### 2.1.2 *Electromagnetic (stripline)*

The electromagnetic pick-up is a transmission line (stripline) that couples to the transverse electromagnetic (TEM) field of the beam. The transmission line is formed between the stripline and the wall of the vacuum chamber and is excited by the beam only at the gaps on either end of the stripline, where a longitudinal field occurs. Figure 4 shows the layout of such an electromagnetic stripline electrode.

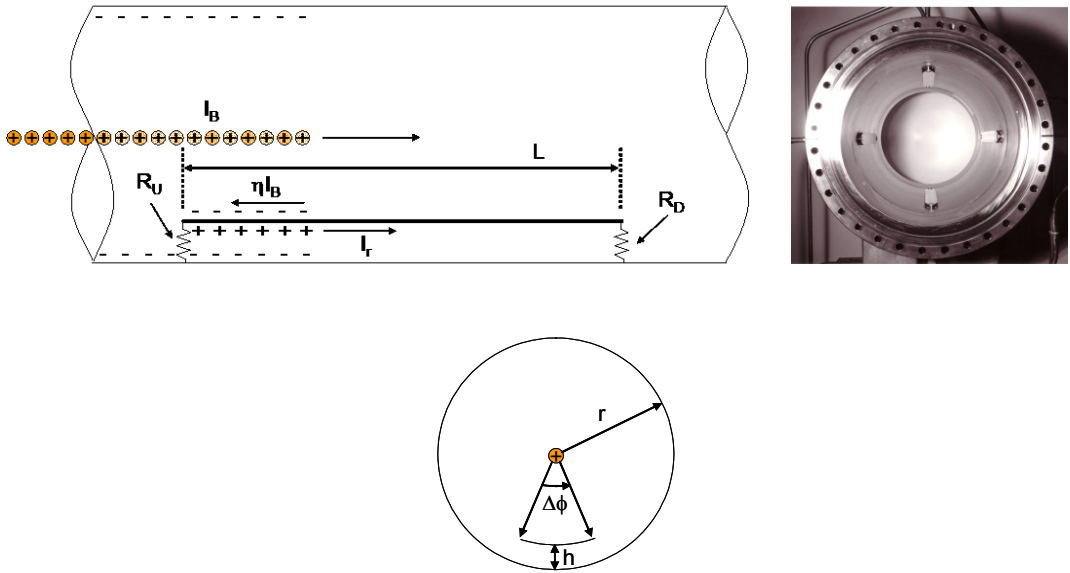

**Fig. 4:** Schematic and photo of an electromagnetic stripline pick-up. $r$ = chamber radius; $h$ = stripline gap height; $w$ = stripline width; $L$ = stripline length; $\Delta\phi$ = angle subtended by the electrode; $\eta = \Delta\phi/(2\pi)$ = fraction of azimuth subtended by electrode; $I_B$ = beam current; $R_U/R_D$ = upstream/downstream termination resistance; $Z_0$ = characteristic impedance of the stripline.

Consider a bunch travelling from left to right (upstream to downstream). While it is over the upstream port there is a voltage $V_r$ across $R_U$, causing a voltage wave of that amplitude to be launched to the right. The stripline forms a transmission line with the wall of the vacuum chamber of characteristic impedance $Z_0$. The voltage wave is therefore accompanied by a rightward-travelling current wave of amplitude $I_r = V_r/Z_0$. This current flows along the bottom surface of the electrode whilst an equal and opposite current flows along the chamber wall. In addition, an image current of amplitude $\eta I_B$ travels along the top surface of the electrode. The voltage $V_r$ across $R_U$ can therefore be expressed as

$$V_r = (-I_r + \eta I_B) R_U = \eta I_B \frac{R_U Z_0}{R_U + Z_0} \quad \Rightarrow \quad V_r = \tfrac{1}{2} \eta I_B Z_0 \tag{2}$$

for a matched stripline ($R_U = Z_0$).

When the beam is over the downstream port it produces a voltage $-V_r = -\frac{1}{2}\, \eta\, I_B Z_0$ across $R_D$ in the same way as it produced a voltage $+V_r$ across $R_U$. This launches a leftward-travelling wave of the same magnitude, but different sign to the rightward-travelling wave, which propagates along the transmission line formed by the stripline and the chamber wall, and will produce an inverted signal upon arrival at the upstream port a time $L/c$ later. The final signal observed at the upstream port will therefore be a bipolar pulse with the maxima separated by $2L/c$ (see Fig. 5(a)).

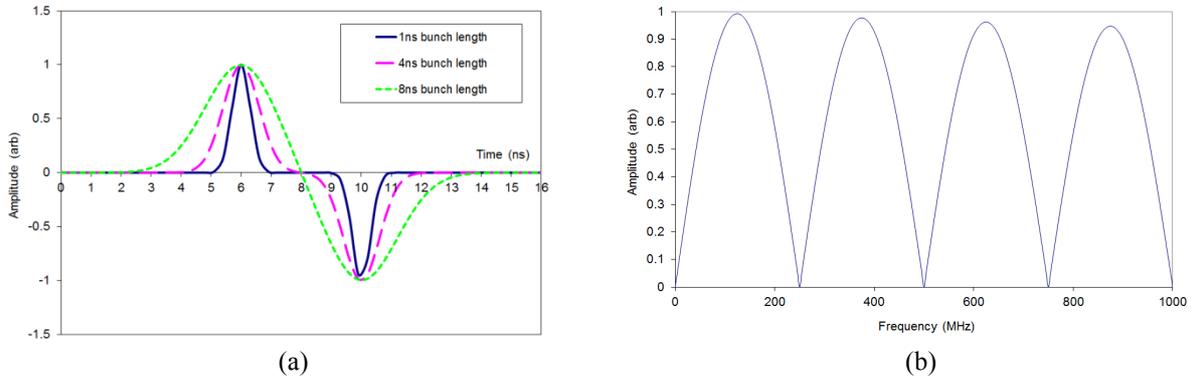

(a)            (b)

**Fig. 5:** (a) Time response and (b) frequency response of a 60 cm-long electromagnetic stripline pick-up

When the radio frequency (RF) wavelength of the beam is equal to multiples of $2L$, the reflection and the signal from the next bunch will cancel and there will be no net signal from the stripline. A maximum in the frequency response will be observed when $L$ is a quarter of an RF period, and hence the stripline pick-up length is usually chosen accordingly. The full frequency response of a 60 cm-long stripline is shown in Fig. 5(b) and has a lobe structure, with the minima located at multiples of $c/(2L)$.

For a relativistic beam the voltage due to the beam passing the downstream port is produced at the same time as the rightward-travelling wave propagating between the stripline and the wall arrives at the downstream port. The two equal and opposite voltages therefore cancel producing no net signal at the downstream port. The electromagnetic stripline pick-up is therefore said to be 'directional', i.e. a signal is only observed on the upstream port with respect to the beam direction. These pick-ups are therefore used in all locations where there are two counter-rotating beams in the same vacuum chamber. Due to imperfections in the stripline and feedthrough impedance matching, the best directivity one can hope to obtain for a real stripline is generally around 25–30 dB (i.e. the voltage signal of one beam with respect to the other is attenuated by a factor between 18–32).

### 2.1.3  *Resonant cavity*

Resonant structures, e.g. 'pill-box' or rectangular cavities, coaxial resonators, and more complex waveguide-loaded resonators, have become very popular to fulfil the high-resolution, single-pass beam position monitoring demands of next-generation, high-energy, linear accelerators [3, 4], or for driving a SASE-FEL beam-line [5].

These are constructed to exploit the fact that an off-centre beam excites a dipole mode ($TM_{110}$) in the cavity, with the amplitude of excitation almost linearly dependent on the off-axis displacement of the beam (Fig. 6). This dipole mode has a slightly different frequency from the main monopole mode ($TM_{010}$) of the cavity, which allows the processing electronics to select only the frequency of interest (dipole $TM_{110}$) and so suppress much of the large, unwanted, intensity-related signal (monopole $TM_{010}$).

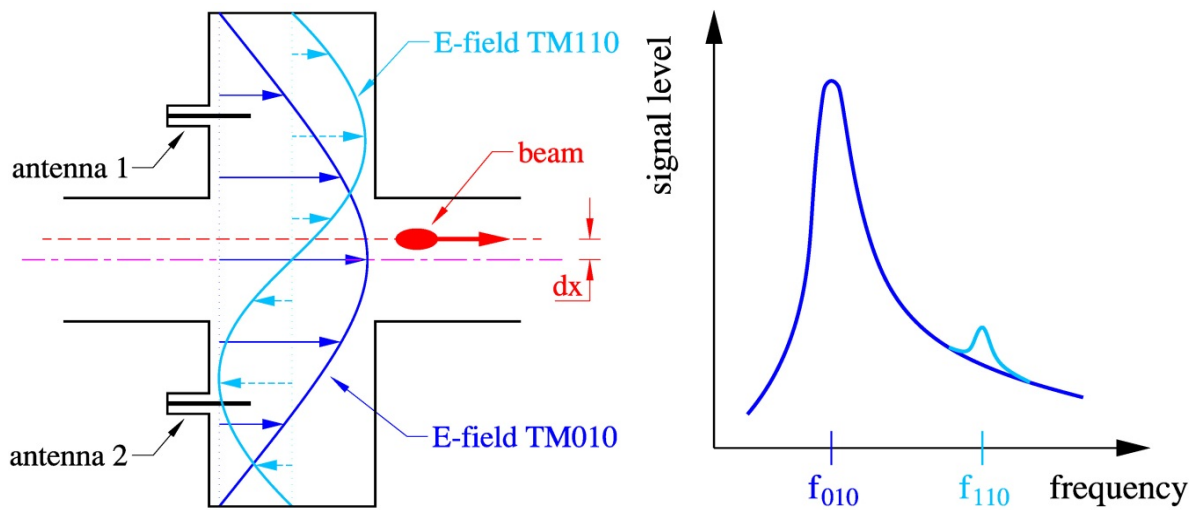

**Fig. 6:** Principle and frequency response of a cavity BPM

Nevertheless, even with this frequency difference, the presence of the fundamental $TM_{010}$ monopole mode still adds a strong common-mode component to the dipole-mode position signal, limiting the performance of the cavity BPM. Rather than picking off the signal from the cavity using four symmetrically arranged pin antennas it is therefore preferable to couple to the cavity using waveguides. Selecting the width of the waveguides such that they have a cut-off frequency above the $TM_{010}$ monopole mode results in a very efficient, internal high-pass filter and makes the cavity BPM quasi 'common-mode free'. Such waveguide-loaded rectangular resonators (Fig. 7) have demonstrated a world-record resolution of 8.7 nm at the ATF2 final focus test beam-line [6].

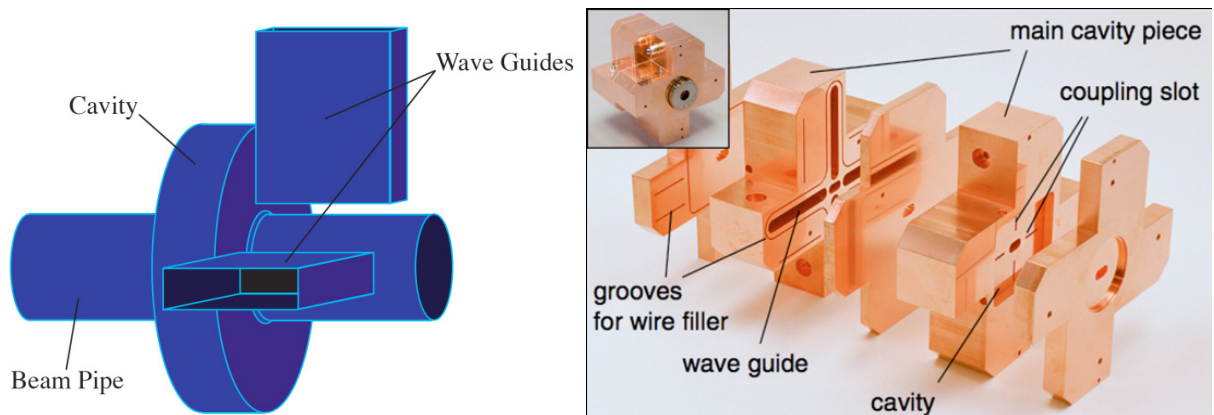

**Fig. 7:** Principle and example of a waveguide coupled cavity pick-up (ATF2, KEK, Japan)

## 2.2 Beam position acquisition systems

Once the signals from the opposite electrodes of a pick-up have been obtained, the next step is to convert these signals into a meaningful beam position. The first thing to do is to normalize the position, i.e. to make it independent of the signal amplitude (i.e. beam intensity). This is generally done using one of three algorithms, whose response curves can be seen in Fig. 8.

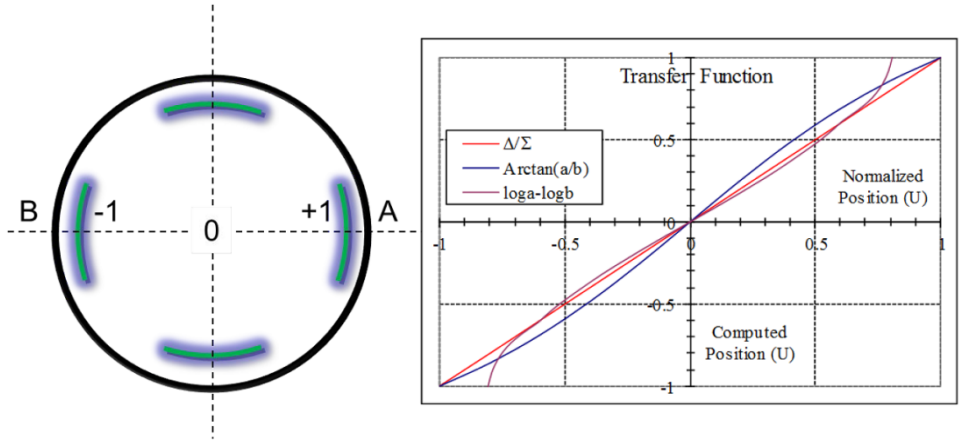

**Fig. 8:** Transfer functions for three commonly used position normalization algorithms

### 2.2.1 Difference over sum (Δ/Σ)

The sum and difference can be obtained either directly from a cavity BPM, or for the other pick-up types using a 0°/180° passive hybrid, a differential amplifier or calculated by software (after digitizing), to give

$$\text{Normalized position} = \frac{A-B}{A+B}. \qquad (3)$$

The transfer function of this algorithm can be seen to be highly linear.

### 2.2.2 Logarithmic ratio

The two input signals are converted into their logarithmic counterparts and subtracted. In practice this is done using logarithmic amplifiers followed by a differential amplifier. This gives

$$\text{Normalized position} = \text{Log}(A) - \text{Log}(B) \left( = \text{Log}\left[\frac{A}{B}\right] \right) \qquad (4)$$

whose response curve is seen to be an reversed 'S' shape, which becomes highly non-linear when exceeding 70% of the normalized aperture.

### 2.2.3 Amplitude to phase

The two input signals are converted by a 90° passive hybrid into signals of equal amplitude but varying phase, with the position dependence of this phase given by

$$\text{Normalized position} = \varphi = 2 \times \arctan(A/B). \qquad (5)$$

Here the transfer function again deviates from the linear in an 'S' form, but does not diverge for large excursions. In addition, the gradient is larger around zero, making it more sensitive towards the middle of the pick-up. A variation on the amplitude-to-phase algorithm is amplitude-to-time conversion, the technique used for the beam position system of the LHC [7].

The type of algorithm to be used will depend on the choice of processing electronics. In all cases the non-linearity is taken into account by calibration circuits and correction algorithms. A summary of commonly used beam position acquisition systems is given in Fig. 9. Here we will only briefly mention the various families in passing, but detailed descriptions along with the advantages and disadvantages of each system can be found in [8] (Fig. 9).

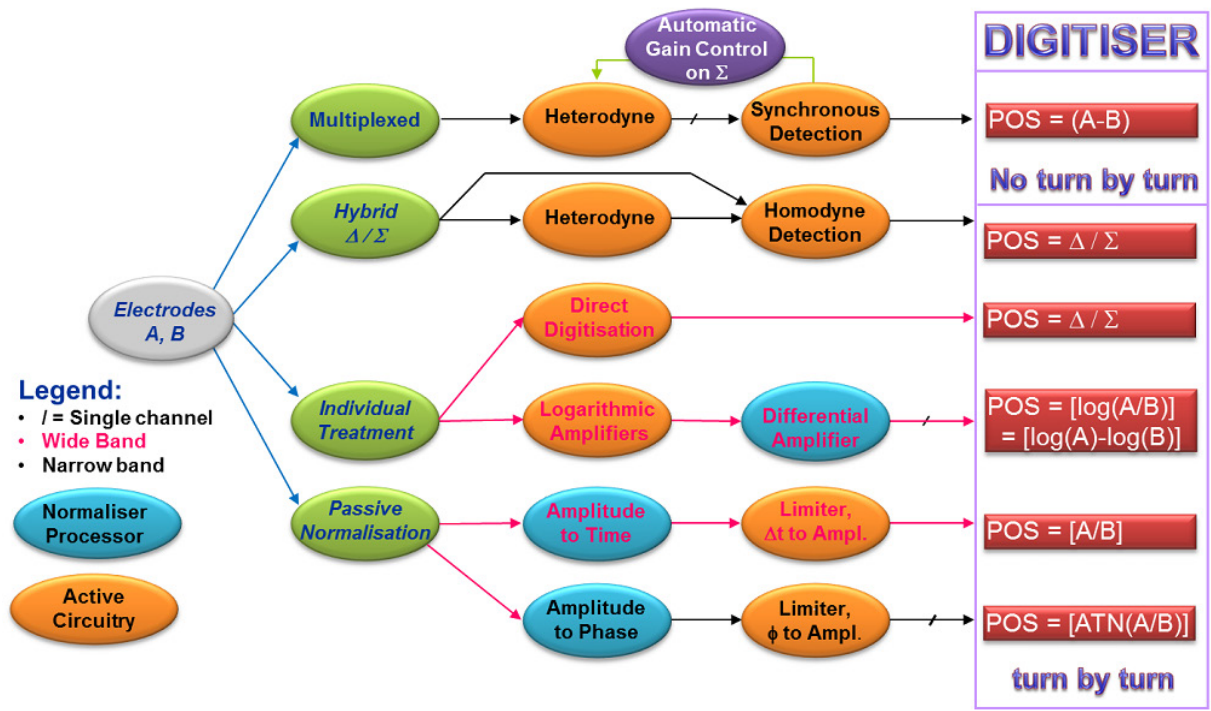

**Fig. 9:** Schematic representation of the various beam position processing families

### 2.2.3.1 MPX (multiplexed)

Each of the BPM electrodes is multiplexed in turn onto the same, single electronics acquisition chain. This eliminates channel-to-channel variations, but since the switching is generally quite slow such an acquisition only tends to be used in circulating machines where the average orbit is of main importance.

### 2.2.3.2 Hybrid (sigma and delta)

Here a 0°/180° passive hybrid is usually used to obtain the sum ($\Sigma$) and difference ($\Delta$) signal from the two electrodes. The position (or ratio of the sum and difference signals) can then be obtained in many ways including direct digitization, homodyne detection (mixing the sum and delta signals with the sum signal itself), or heterodyne detection (mixing sum and delta signals with an external reference).

### 2.2.3.3 Individual treatment

In this case each electrode is treated separately, but in parallel. The acquisition can either consist of directly digitizing each signal or using logarithmic amplifiers as outlined above. The disadvantage of this method is that it requires two (or four depending on the pick-up orientation) very well-matched chains of electronics, since the combination of the signals to obtain a position is performed at the very end of the chain.

### 2.2.3.4 Passive normalization

Here the amplitude difference (i.e. position information) in the input signals is directly converted into a phase or time difference. Intensity information is lost in this procedure, but the result is a varying phase or time that is directly proportional to the position.

## 2.2.4 Read-out electronics

The read-out system interfaces the BPM pickup to the accelerator data acquisition (control) system. This requires signal conditioning, normalization, and linearization of the BPM signals with conversion to a digital format somewhere along this chain in order to ultimately provide a time-stamped beam position.

Modern BPM read-out electronics are typically based on the individual treatment of the electrode signals using frequency domain signal processing techniques [9]. These techniques were developed for the telecommunications market and make use of the high-frequency and high-resolution analogue-to-digital converters (ADCs) that are now readily available. In such schemes, bandpass filters in the analog section convert the BPM signals into sinewave-like signal bursts for waveform sampling and processing in the following digital electronics. Microwave and RF analogue components, 12–16 bit pipeline ADCs, field programmable gate arrays (FPGAs), and clock distribution chips with sub-picosecond jitter are some of the key hardware elements. Figure 10 illustrates a typical electronics arrangement for a broadband BPM pickup, with only one of the four channels shown. For cavity BPMs the schematic is similar.

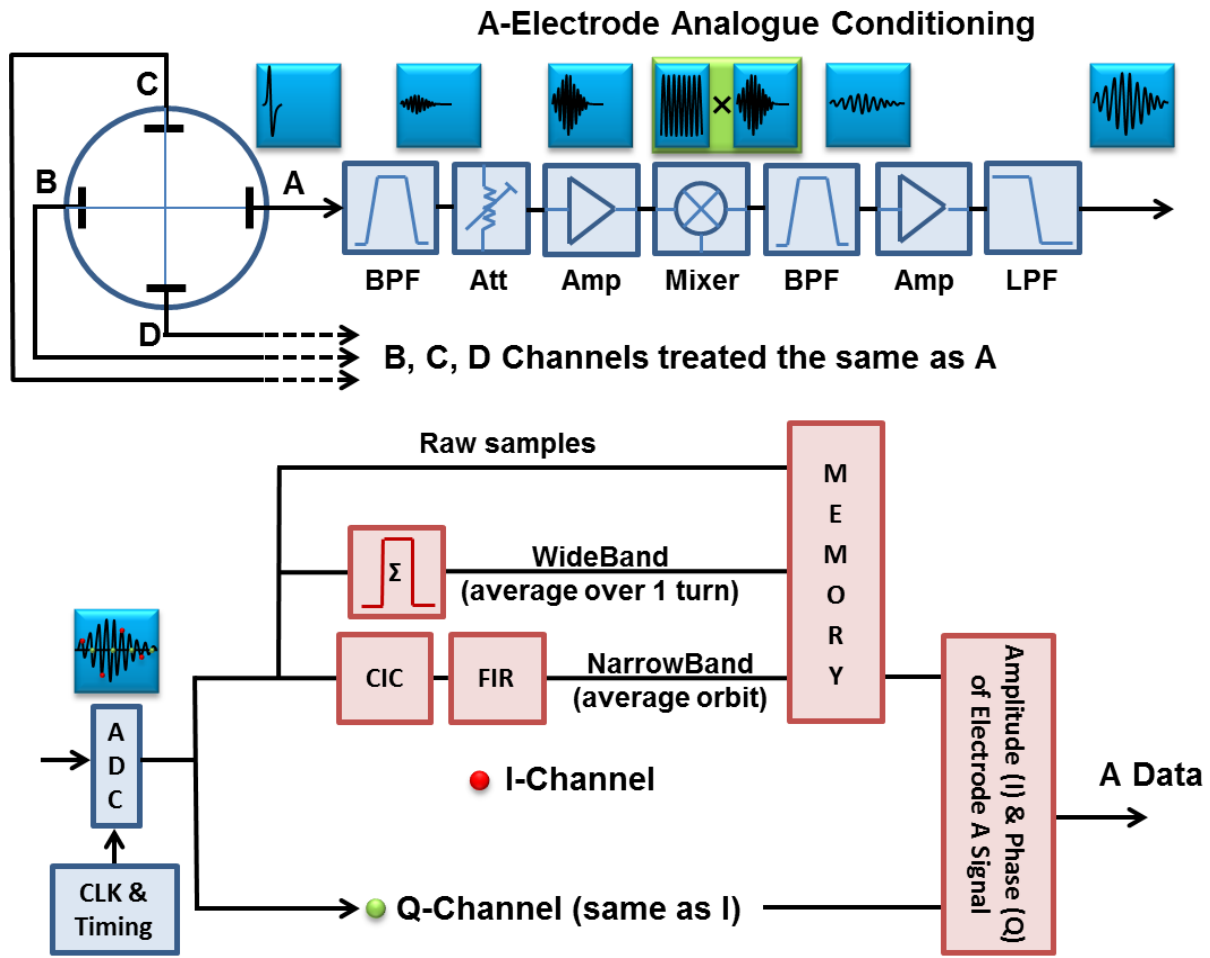

**Fig. 10:** Key elements of modern BPM read-out electronics

The analogue chain, consisting of bandpass filters, amplifiers, and typically a frequency down-conversion stage prepare the electrode signal for sampling by an ADC. In order to reconstruct the input sine wave the ADC is either clocked at some sub-harmonic of the accelerator radio frequency or with an external clock, a numerically controlled oscillator (NCO). This clock is typically chosen to give a sampling at four times the frequency of the input sine wave or to under-sample the input sine wave by a multiple of four. This allows I/Q demodulation to be carried out in the digital domain. In-phase/quadrature (I/Q) demodulation is nothing more than sampling at some fourth multiple or sub-harmonic of the input frequency. The knowledge that the sampled points are then all 90° apart on the sine wave allows for easy computation in the digital domain without the need for sine and cosine look-up tables. If the frequency is correct, and the phase is locked, then all the quadrature samples are zero and only the in-phase samples need to be considered, giving directly the amplitude of the sine wave. These data can then be treated either in their raw form for bunch-to-bunch measurements, in a wideband form for turn-by-turn measurements or in narrowband for orbit measurements. The narrowband orbit data is typically always available online from such systems, with the wideband and raw data available on request for a limited number of turns or bunches, respectively. By producing the sum and difference of the amplitudes from opposite electrodes and applying calibration and linearization factors this data can then be converted into a meaningful beam position.

## 3    Beam current and intensity measurement

The measurement of beam current or bunch intensity is one of the most basic and important measurements performed at any accelerator, used to optimize its operation and to calibrate physics experiments. For determining the time evolution, relative measurements are often sufficient, relying only on a good resolution from a single detector. However, when comparing the intensity read-out from several devices, such as for transfer line or injection optimization, or when using the system to cross-calibrate other devices, a precise system calibration is mandatory.

### 3.1    The beam current transformer

The beam intensity is usually measured by means of a beam current transformer (BCT). This name originates from the fact that in such a device the beam can be considered as the primary, one-turn winding of a transformer, with its equivalent current 'transformed' to the secondary winding output (see Fig. 11).

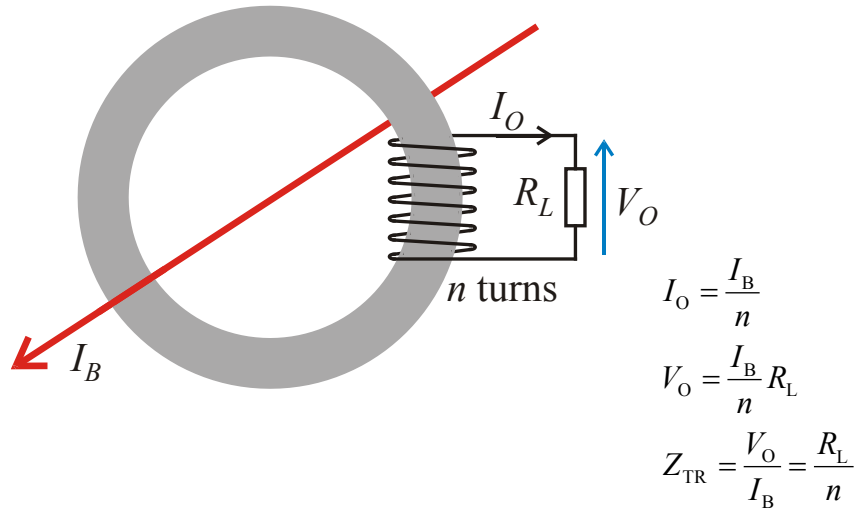

$$I_O = \frac{I_B}{n}$$

$$V_O = \frac{I_B}{n} R_L$$

$$Z_{TR} = \frac{V_O}{I_B} = \frac{R_L}{n}$$

**Fig. 11:** The principle of a beam current transformer

The transformer is based on two principles: first, that an electric current produces a magnetic field, and second that a changing magnetic field within a loaded coil of wire induces a current in the coil (electromagnetic induction). Changing the current in the primary coil changes the magnetic flux that is developed, with the changing magnetic flux inducing a current in the secondary coil.

The BCT relies on a toroid made of ferromagnetic material to capture the magnetic field, passing it through an *n*-turn secondary winding loaded with a resistance $R_L$. A beam current $I_B$ passing through the toroid induces a current $I_O = I_B/n$ in the secondary $n$-turn winding. This current is converted into an output voltage $V_O = I_O \times R_L$. The BCT transfer impedance, i.e. the factor determining the output voltage corresponding to a given beam current, can be defined as $Z_{TR} = R_L/n$.

A perfect BCT would reproduce the temporal distribution of the beam current shape with no distortion and amplitudes described with the above equations. Each real BCT has a limited bandwidth, which can be characterized by the low and high cut-off frequencies, $f_{cL}$ and $f_{cH}$, respectively. Typically they are chosen according to the spectral content of the beam to minimize the distortion of the beam pulses.

The low cut-off originates from the fact that the impedance of the secondary winding decreases for low frequencies. The secondary winding inductance is given by $L_S = A_L \times n^2$, where $A_L$ is the toroid inductance constant having the unit nH/turn². $A_L$ depends on the toroid material and its geometry, in particular the length to cross-section ratio. At the cut-off frequency $f_{cL}$ the inductive reactance $X_L = 2\pi f_{cL} L_S$ of the secondary equals the load $R_L$, giving the BCT low cut-off frequency

$$f_{cL} = \frac{R_L}{2\pi A_L n^2} . \qquad (6)$$

For modern, high-permeability materials $A_L$ is in the order of 10 000 nH/turn². For a BCT with such a toroid, a secondary winding with $n = 30$ turns and load $R_L = 10\ \Omega$, the low cut-off frequency would be some 180 Hz.

The fact that the BCT has a low-frequency cut-off implies that the output signal will have no DC component, which results in a distortion of the signal. For long pulses with a low duty cycle (e.g. from a linac) the distortion is called 'droop' and is sketched in Fig. 12(a). The amount of droop at the end of the pulse depends on the ratio of the pulse length $\tau_p$ and the time constant $\tau_{cL} = (2\pi f_{cL})^{-1}$. For $\tau_p \ll \tau_{cL}$ the dependence is linear. Pulse trains of short bunches suffer from a phenomenon known as 'baseline shift', depicted in Fig. 12(b), where the steady-state integral of the BCT output signal is zero. This is typical for circular accelerators and is also a consequence of the low-frequency response and lack of DC component. Special compensation techniques are therefore often required to restore the missing DC response and are known as 'baseline restoration' techniques. In the past this was done electronically, often requiring complicated switching circuitry and timing, whilst nowadays it can be efficiently performed in the digital domain.

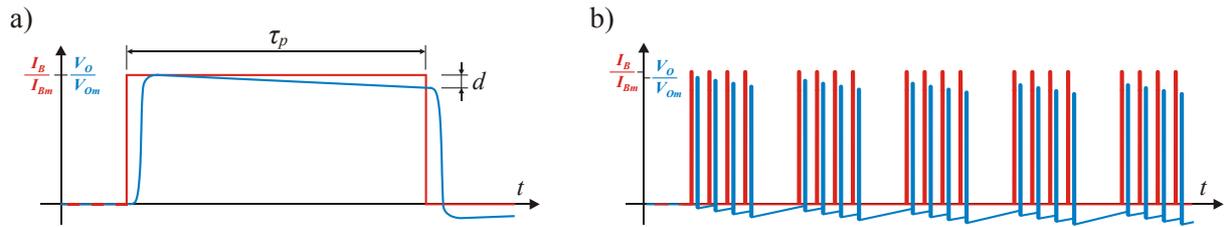

**Fig. 12:** Influence of the limited BCT low cut-off frequency: (a) signal droop; (b) baseline shift

If the BCT frequency characteristic below the low cut-off frequency $f_{cL}$ is defined by the first-order high-pass of the secondary winding inductance and its load, it can be compensated by an analogue circuit. This can be achieved using an amplifier having more gain for these low frequencies, as sketched in Fig. 13. Such a circuit can be built using an operational amplifier with the feedback circuit shown. For low frequencies its gain is given by $g_L = R_1 / R_2 + 1$, while for high frequencies the gain is given by $g_H = (R_3 \| R_1)/R_2 + 1$, where $(R_3 \| R_1)$ is the resistance of $R_3$ and $R_1$ in parallel. Capacitance $C$ is chosen according to $f_{cL}$ in order to achieve a flat overall frequency response. The price to pay for this higher low-frequency gain of the amplifier is deterioration in the noise performance for the compensated frequency band.

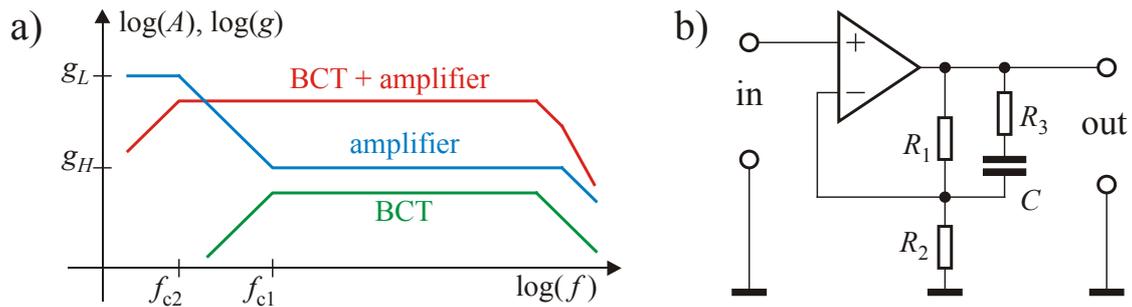

**Fig. 13:** (a) Simplified frequency characteristics of a BCT with a correcting amplifier; (b) an implementation example using an operational amplifier.

The high-frequency response of a BCT depends on many factors, making it very difficult to calculate and even to simulate with modern electromagnetic finite-element simulators. In practice the high-frequency response is therefore still often optimized on prototypes. It depends on the toroid material, toroid size, toroid winding load, the ceramic insert, and winding technique used, but always decreases with the number of turns. BCTs optimized for bandwidth can cover a range of 6–7 decades in frequency, with the low cut-off typically ranging from 100 Hz to 1 kHz and the high cut-off from 100 MHz to 1 GHz.

An example of BCT construction is sketched in Fig. 14. In order for the transformer to interact with the magnetic field of the beam it has to be placed over a ceramic gap in the vacuum chamber. To keep the impedance seen by the beam as low as possible an RF bypass (either a thin metallic coating or external capacitors on the ceramic) is required for the very-high-frequency wall current components. In addition, to keep vacuum chamber continuity, the transformer is enclosed in a housing connected galvanically to the beam pipe.

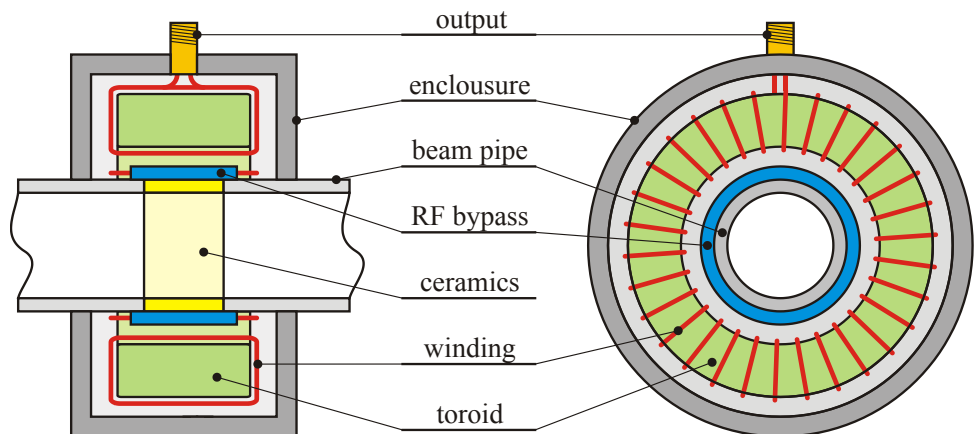

**Fig. 14:** The beam current transformer

An example of a wide-bandwidth BCT system is the one installed on the CERN-SPS [10], with a bandwidth from 200 Hz to 1 GHz. Such a bandwidth is obtained by using a ferromagnetic core wound of high-permeability amorphous metal tape. With this system operators can observe the bunch-by-bunch intensity evolution of beams throughout the SPS acceleration cycle. In order to obtain the total charge in each bunch, fast integrators are used, capable of working at repetition frequencies of up to 40 MHz. The integrators are based on a dedicated chip, developed primarily for capturing photomultiplier signals in the LHC-b experiment [11]. A schematic of the integration principle and the resulting signals as measured in the CERN-SPS are shown in Fig. 15. The chip works using two integrators operating in parallel. As one integrates the other is discharged, with the output switched from one to the other on each clock cycle. The resulting integrated signals are directly digitized, with all integrator linearization and baseline restitution performed in the digital domain.

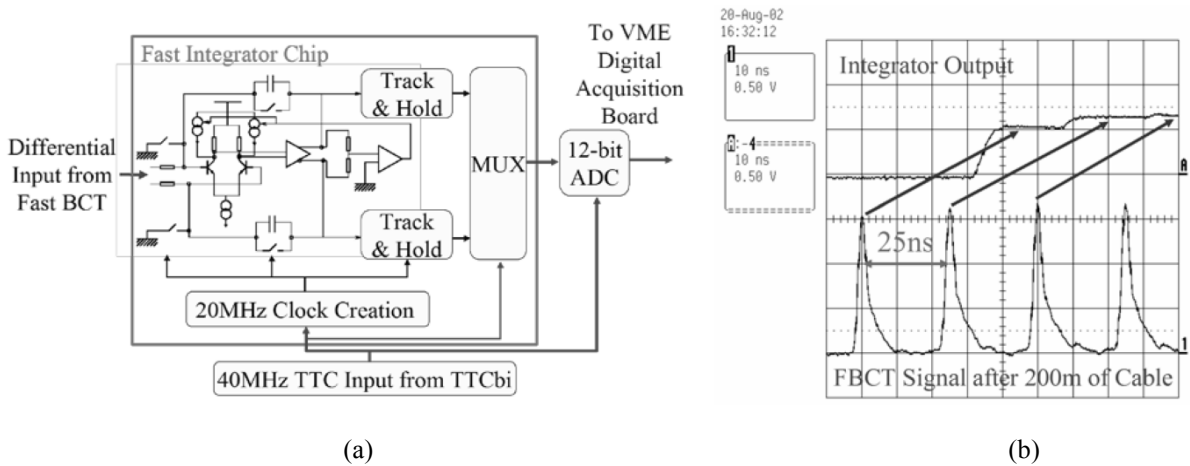

(a)                      (b)

**Fig. 15:** (a) The integrator of the CERN-SPS BCT; (b) a beam measurement example

## 3.2 The DC beam current transformer

In storage rings and accelerators with cycle times of several seconds, a DC beam transformer can be used to measure the total current. Such an instrument was developed for the CERN Intersecting Storage Rings (CERN-ISR) in the early 1980s, the first machine to sustain beams for hours [12]. A DC transformer is based on a pair of matched, toroidal, ferromagnetic cores, which are driven into saturation by a modulation current at frequencies of up to a few kHz. The principle of operation is shown in Fig. 16, and makes use of the hysteresis loop of the toroid.

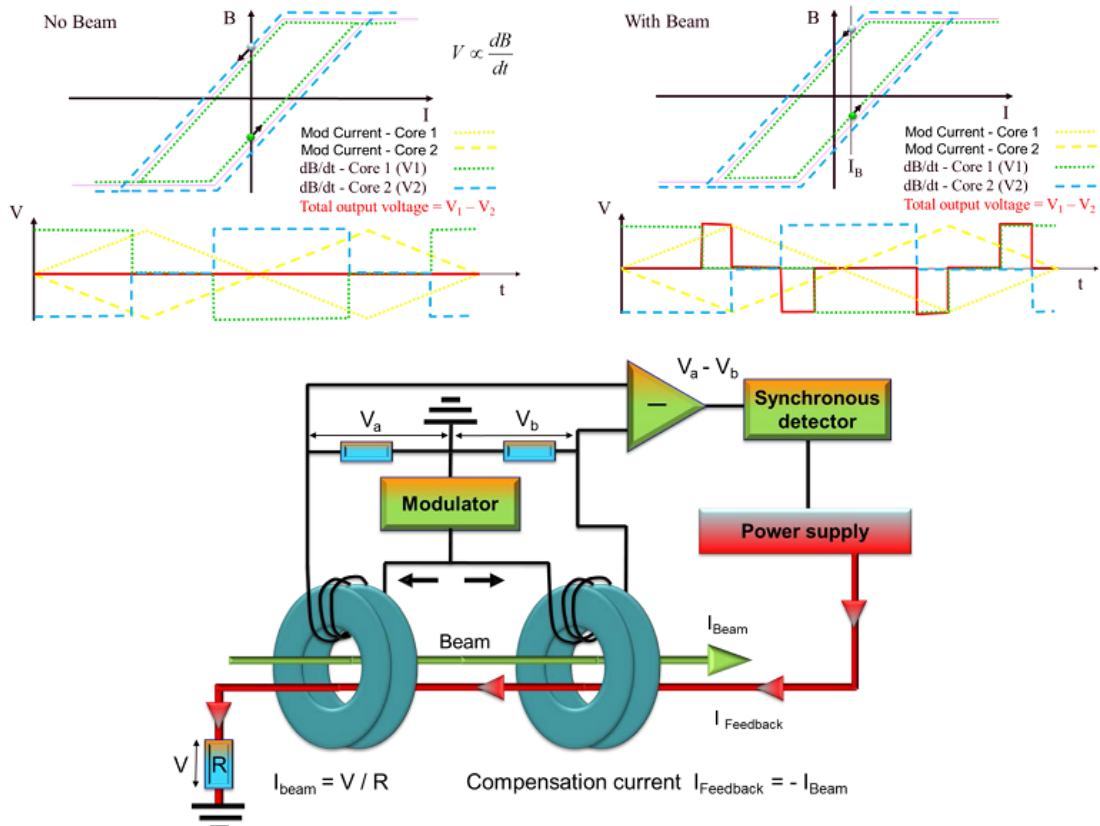

**Fig. 16:** The principle and schematic of a DC beam current transformer

If an equal but opposite modulation current (the triangular waveforms in Fig. 16) is applied to both cores with the beam not present, then the voltage induced in the detection windings on each core will be equal but opposite. When, however, there is a beam current $I_B$ present, the starting point in the hysteresis loop for zero modulation current is offset due to the static magnetic field generated by the beam current. Since the modulation is in opposite directions in each toroid, the time spent in saturation will be different for the two branches of the hysteresis loop. This results in the generation of voltage pulses at twice the modulation frequency when the induced voltage in the detection windings on each core is combined. The demodulation of this signal gives a train of pulses, with the width of each pulse being a direct measure of the beam current, i.e. by how much the hysteresis curves are offset.

In the 'zero flux detector' implementation of the DC beam transformer, the result of the demodulation is fed back into a compensating current loop (see Fig. 16). Once the compensation current and the beam current are identical the net static magnetic field seen by the toroids is zero (hence zero flux) and the output from the demodulator is also zero. The beam current can then be obtained by simply measuring the voltage produced by this compensation current across a known resistor.

For modern DC transformers such a zero flux detector is used to compensate the droop of the simple beam current transformer described in Section 3.1. This significantly increases the bandwidth of the system, allowing measurement from DC to a few MHz.

## 4  Diagnostics of transverse beam motion

The instrumentation used to observe transverse beam motion is very important for the efficient operation of any circular accelerator [13]. There are three main parameters that can be measured using such diagnostics, namely the betatron tune, chromaticity, and coupling, all of which are discussed in detail below.

### 4.1  Tune measurement

Measuring and optimizing the betatron tune through the whole operational cycle of a circular accelerator is one of the most important and basic control room tasks, and strongly influences the beam lifetime and quality. Since the betatron oscillations are typically only detected at one point in the machine the measurement gives the fractional part of the tune, which is the quantity of interest for beam optimization. Separate measurements are normally performed in order to obtain the tune in both the horizontal and vertical planes. The quality of the tune measurement is very important for measuring other crucial beam parameters, such as chromaticity and betatron coupling, which are traditionally derived from tune measurements.

Tune measurement systems can be divided into two main categories of system (see Fig. 17), those based on turn-by-turn BPM measurements followed by special tune signal processing and those constructed as dedicated systems optimized for detecting only the oscillations of the beam. The 'observation method' can either be wideband, i.e. yielding a full response function over the frequency range of interest, or narrowband, detecting only the response at a given frequency. In order to be able to observe an oscillation an appropriate excitation has to be applied to the beam, the 'operation mode'.

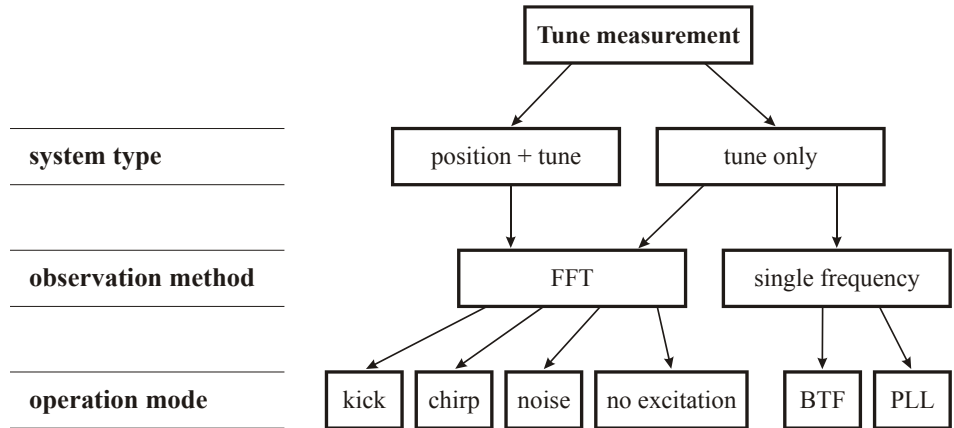

**Fig. 17:** Categories of tune measurement systems

*4.1.1  Tune measurement systems*

A tune measurement system can be based upon turn-by-turn readings from a regular beam position acquisition system, such as those described in Section 2. Signal processing of these signals typically results in spectra showing the oscillation frequencies present in the transverse beam motion. Such a universal system is, in many cases, sufficient, especially for machines that can tolerate relatively large beam excitation (such as electron machines, where radiation damping counters any emittance blow-up introduced by exciting the beam for a short period). Its main limitation is sensitivity, originating from the fact that, in most cases, the BPM system is optimized for accurate position measurements over a large dynamic range in both position and beam intensity. For example, if the system has to measure positions over ±10 mm, then to detect oscillations with 1 μm amplitude, a dynamic range of 10 000 is required in turn-by-turn mode. This is, in practice, difficult to achieve, with position measurements typically averaged over many hundreds of turns to achieve such a resolution. This can be improved significantly by building dedicated systems optimized for beam oscillation detection. In such systems the static beam position is rejected at a very early stage with only the oscillation signal used for further processing. An example is the so-called BaseBand Tune (BBQ) system, based on a direct diode detection method initially designed for the LHC [14]. This allows a reliable tune measurement with micrometre beam oscillations, which in some machines is always present without the need for additional, explicit excitation.

In this technique the electrode signals from a beam position pick-up undergo envelope demodulation with diode peak detectors, which can be considered as a simple, fast, sample-and-hold circuit (see Fig. 18). Most of the pick-up signal does not change from one turn to another and gets converted to DC, while the small beam oscillation modulation is kept as ripples on this DC offset. A series capacitor blocks the large DC component related to the intensity and static beam orbit, while allowing the small ripples to pass and be amplified. As the beam oscillation modulation from opposite pick-up electrodes is of opposite phase, enhancement of the signal is possible by subtraction using a differential amplifier, with all common-mode interference and remaining intensity and orbit signals further suppressed. The tune range of interest typically corresponds to half the machine revolution frequency, ranging from tens of kHz to MHz depending on the size of the accelerator. The beam oscillations of importance after the diode peak detectors are therefore of relatively low frequency and as such can be easily filtered, amplified, and digitized with high-resolution ADCs. Further details on the direct diode detection technique along with some adventures encountered during its development can be found in [15], which also includes a list of references related to the subject.

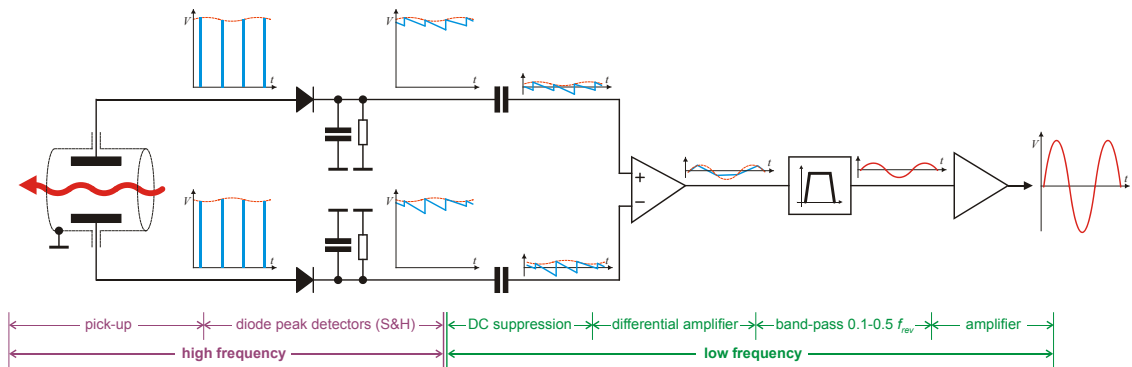

**Fig. 18:** Principle of the baseband tune measurement system using direct diode detection

### 4.1.2 *Tune measurement observation methods*

Narrowband tune observation was popular in the past when high-frequency digitization and processing of RF signals was expensive and difficult to achieve with enough precision. In such cases most of the signal processing was performed in the analogue domain with RF mixers and filters, with only the resulting low-frequency content digitized and analysed. They always involve locking the detection frequency to the excitation frequency, and measuring the beam response at that particular frequency. Beam transfer function (BTF) measurements and phase-locked loop (PLL) measurements are the two main examples of narrowband tune observation and will be discussed further in the next section.

With the recent advance in high-resolution, high-frequency digital electronics, most modern tune observation is performed wideband. The advantage of wideband tune observation is that it yields a complete spectrum of the transverse beam motion. In addition to tune measurement this then allows detection of beam instabilities and abnormal sources of beam excitation, typically originating from malfunctioning of an accelerator system, such as the RF system or magnet power supplies. Wideband observation is based on a fast sampling of the input signal followed by analysis using fast Fourier transform (FFT). In the simplest case the tune is localized in the beam spectrum as the frequency corresponding to the highest spectral peak. In practice, however, the beam spectrum is often much more complicated. Use of wideband observation in such cases is a big asset, as it allows tune determination using algorithms of practically unlimited complexity, something that is simply not possible in single-frequency systems.

### 4.1.3 *Tune measurement operation modes*

In a circular accelerator each particle in the beam undergoes a transverse harmonic motion around the ideal orbit defined by the magnetic lattice. However, the resonant frequency (tune) of each particle is slightly different, resulting in an incoherent motion. To an external observer (such as a beam position pick-up), which looks just at the centre of mass of all these moving charges, this yields a constant signal that is directly proportional to the total intensity and the average position. In order to measure the average resonant frequency (tune) and the spread in frequencies (tune spread) the movement of the particles need to be synchronized, something that is achieved using externally applied excitation.

#### 4.1.3.1 *Excitation techniques for FFT-based measurements*

A classical tune measurement consists of applying a transverse kick to the beam, typically by discharging a capacitor bank to generate a large deflecting current in a kicker magnet. An example of such a measurement is shown in Fig. 19.

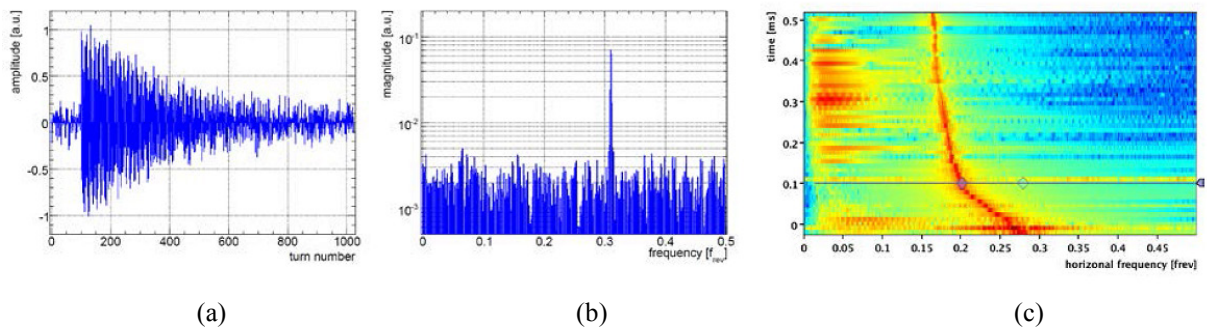

(a) (b) (c)

**Fig. 19:** (a) Proton beam response to single kick excitation with (b) corresponding magnitude spectrum. (c) Tune measurement example from the CERN Booster Synchrotron with kicks every 10 ms.

The kick has the effect of offsetting all the particles by the same amount from the reference orbit, from which they then continue to perform their individual harmonic motion. The average frequency of this motion is what is known as the tune. The decay in the amplitude of the oscillation, as seen in the pick-up signal, is a result of the slightly different frequencies of the individual particles. After a while this results in the motion becoming incoherent, observed as a damping of the oscillation signal. What should not be forgotten is that in a hadron machine all particles still undergo the original harmonic motion, which is now higher in amplitude due to the applied kick. This results in an increase of the effective beam size and hence emittance. In order to follow the tune evolution along an acceleration cycle the kicks can be repeated, giving a full time-resolved beam spectrum as shown in Fig. 19.

Another commonly used excitation technique is the so-called 'chirp'. This involves sweeping the applied excitation frequency over a certain range where the tune is expected to sit. The name chirp comes from the fact that if listened to at audio frequencies such a signal sounds like the chirp of a bird.

The sweep time should be short enough so that the machine tune can be assumed to remain constant during the excitation time. Typically, tune changes are slow compared to the revolution frequency. The sweep range and time can therefore be chosen in such a way that the beam gets coherent kicks over several machine turns when the sweep reaches the right frequency. Beam oscillations of sufficient amplitude can thus be achieved with much smaller peak excitation power than with the single kick method. Chirp excitation trades the excitation strength with the excitation duration, and can often be applied to the beam through a direct input to the transverse feedback system.

Noise excitation, an alternative to chirp excitation, continuously puts power into the beam spectrum at all frequencies. This has the advantage that it does not require any synchronization between the excitation source and the beam signal acquisition. In some machines various unwanted noise sources result in sufficient 'natural' beam oscillations that no additional excitation is required for tune measurement.

*4.1.3.2 Excitation techniques for single-frequency-based measurements*

A beam transfer function (BTF) measurement consists of exciting the beam with a steady sinusoidal signal and measuring the resulting beam motion at this specific frequency. The excitation frequency is then changed and a new measurement made, with this process repeated until the range of interest is covered. Both the amplitude and phase of the resulting beam oscillation can be precisely determined, as the excitation frequency is known. A BTF measurement typically takes quite a long time, during which it must be assumed that all machine conditions remain constant. This is why the BTF method cannot be used for studying dynamic phenomena.

An example of a beam transfer function is shown in Fig. 20. Notice how the phase jumps by 180° as the excitation is applied either side of the central betatron tune frequency. Such a response is typical for any harmonic oscillator.

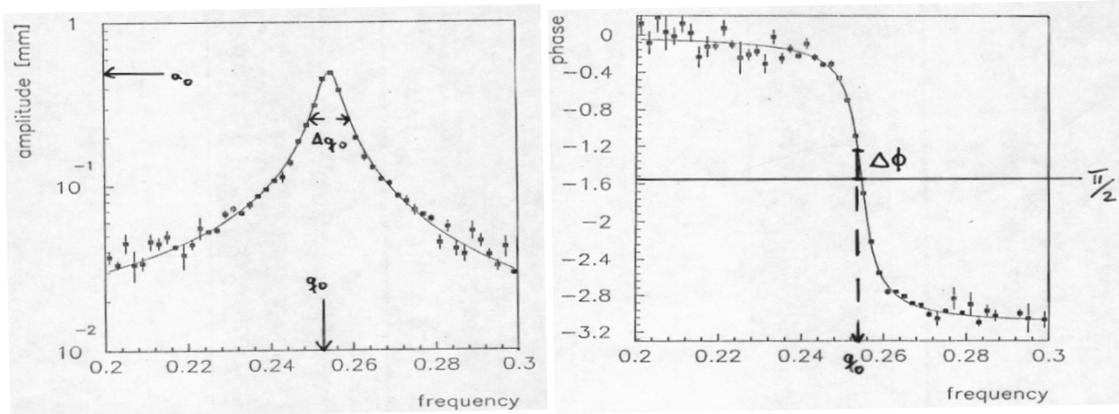

**Fig. 20:** Complete beam transfer function measured on the CERN-LEP

This 180° phase jump is the basis for a PLL tune tracker. Its basic principle is sketched in Fig. 21. A voltage controlled oscillator (VCO) or numerically controlled oscillator (NCO) is used to put a harmonic excitation $A \cdot \cos(\omega t)$ onto the beam. The beam response to this signal is then observed using a position pick-up. The response is of the form $B \cdot \cos(\omega t + \phi)$, where $A$ and $B$ are the amplitude of the excitation and oscillation respectively, $\omega$ is the tune expressed as angular frequency and $\phi$ is the phase difference between the excitation and the observed beam response. In the phase detector both signals are multiplied, resulting in a signal of the form

$$\tfrac{1}{2}AB \cdot \cos(\phi) + \tfrac{1}{2}AB \cdot \cos(2\omega t + \phi) \,, \tag{7}$$

which has a DC component proportional to the cosine of the phase difference. This DC component is therefore zero when the phase difference is 90° which, as can be seen in Fig. 20, is when the amplitude response has a maximum, i.e. at the tune frequency. The aim of the PLL is therefore to 'lock-in' to this 90° phase difference between excitation and observed signal by correcting the VCO frequency until the DC component of the phase detector output is zero. Since the PLL always tries to maintain this 90° phase difference, the VCO frequency tracks any tune changes, resulting in a continuous tune measurement.

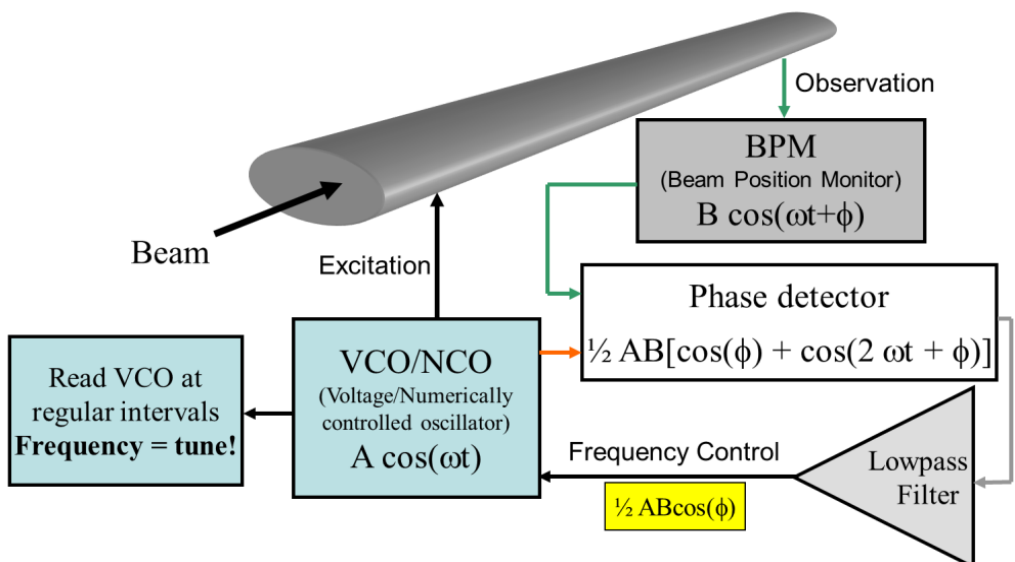

**Fig. 21:** Principle of a PLL tune tracker

In practice, PLL tracker operation is influenced by many parameters, which have to be optimized so that the PLL finds, locks-in, and subsequently tracks the tune changes. The beam spectra and dynamics have to be well understood if the PLL is not to lock or jump to a synchrotron sideband or an interference line. In addition, for hadron machines, the continuous excitation required leads to emittance blow-up.

PLL tune tracking has been used on many machines, and is the basis upon which simultaneous feedback on tune, chromaticity, and coupling was made possible at the Relativistic Heavy Ion Collider (RHIC) at Brookehaven National Laboratory [16].

### 4.2 Chromaticity measurement

For any high-energy synchrotron, the control of chromaticity is very important. If the chromaticity is of the wrong sign (corresponding to positive below the transition energy or negative above it) then the beam quickly becomes unstable due to the head–tail instability. If the chromaticity is too big then the tune spread becomes large and some particles are inevitably lost as they hit resonance lines in tune space. The most common method of measuring the chromaticity of a circular machine is to measure the betatron tune as a function of the beam energy and then to calculate the chromaticity from the resulting gradient. This is usually done by varying the RF frequency, keeping the magnetic field static. The equations of interest are:

$$\Delta Q = (\xi Q)\frac{\Delta p}{p} = Q'\frac{\Delta p}{p} = Q'\gamma_t^2 \frac{\Delta R}{R} = Q'\left(\frac{-\gamma_t^2 \gamma^2}{\gamma^2 - \gamma_t^2}\right)\frac{\Delta f}{f} \quad (8)$$

where $\Delta Q$ is the change in tune, $\Delta p/p$ is the momentum spread (or relative change in momentum), $\Delta R/R$ the relative change in radius, $\Delta f/f$ the relative change in RF frequency, $\gamma$ and $\gamma_t$ the relativistic $\gamma$ and $\gamma$ at transition respectively, and $\xi$ the chromaticity. Please note that the chromaticity is often expressed as $Q' = Q\xi$, where $Q$ is the total betatron tune including the integer part.

In the CERN-SPS, for example, a chromaticity measurement consists of performing a tune measurement for three different RF frequency settings. Instead of noting the exact RF frequency, what is actually measured is the change in closed orbit, from which the relative change in radius can be calculated. These three points are then plotted, with the gradient giving the chromaticity.

In order to obtain continuous chromaticity measurements this technique of RF modulation is combined with one of the tune measurement methods described in the previous section. By tracking the tune during this time and knowing the magnitude of the RF change, the chromaticity can be calculated and measured online.

### 4.3 Coupling measurement

The control of coupling (the degree to which horizontal and vertical betatron motion is linked) is also important for circular accelerators. Excessive coupling will make tune and chromaticity measurements almost impossible, as the information from both planes are mixed-up in the observed signal. A very good and comprehensive summary of linear betatron coupling can be found in [17].

#### *4.3.1 Closest tune approach*

For this method, both betatron tunes are measured during a linear quadrupole power converter ramp that crosses the values of the horizontal and vertical tunes. The remaining separation of the tune traces is a direct measure for the total coupling coefficient $|c|$. A measurement example from the CERN-LEP, using a PLL tune measurement is shown in Fig. 22. In order to ensure that the PLL keeps tracking both tunes, even when they approach each other, the measurements are performed on two different bunches.

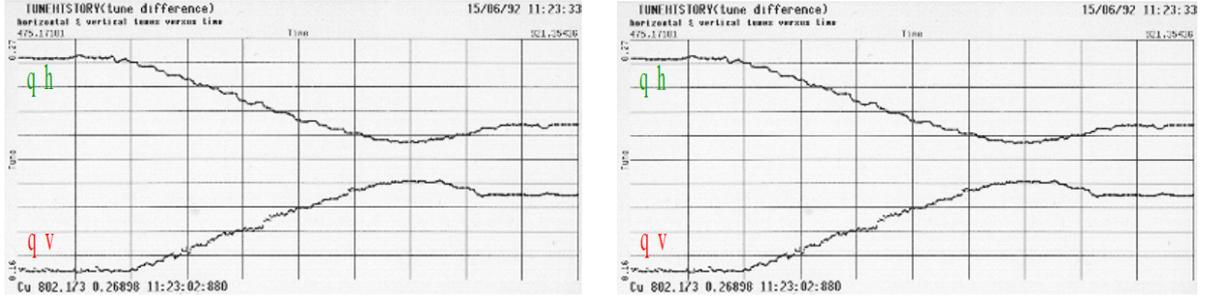

**Fig 22:** LEP coupling measurement

*4.3.2    Kick and PLL methods*

The above method does not allow for diagnostics during machine transitions. A better tool for the measurement of small coupling coefficients, although demanding quite large beam excitations, consists of applying a single kick in one plane and observing the time evolution of the betatron oscillations in both planes. This method is described in [13].

Alternatively, the PLL described in Section 4.1.3.2 can be used in a special configuration to measure both the amplitude and phase of the excitation seen in the other plane. This allows an online coupling measurement with minimal excitation and, in the presence of strong coupling, is essential for robust PLL tune control [18].

## 5    Emittance measurement

The ultimate luminosity of any collider is inversely proportional to the transverse emittance of the colliding beams. Preservation of emittance and hence emittance measurements are therefore of particular importance in the long chain of accelerators and storage rings of big hadron colliders, as the emittance of a hadron bunch is not appreciably reduced through mechanisms such as the radiation damping associated with lepton machines.

In lepton machines, achieving the smallest possible vertical emittance is a common goal for damping rings for linear colliders, for particle factories based on the crab-waist collision scheme, and for light-source storage rings providing photon beams of highest brightness. The measurement of such small emittance beams is therefore very important and is one of the main challenges for modern day beam instrumentalists.

Good explanations of emittance can be found in Refs. [19, 20]. The emittance, which includes about 98% of the beam-particles, can be defined as

$$\varepsilon(98\%) = \frac{beamwidth^2 - \left(\frac{\Delta P}{P} \cdot D_m\right)^2}{\beta_m} = \frac{FWHM^2 - \left(\frac{\Delta P}{P} \cdot D_m\right)^2}{\beta_m} \qquad (9)$$

where *FWHM* is the measured full width at half height (2.35 $\sigma$) of the beam, $\Delta P/P$ the *FWHM* of the momentum spread, $D_m$ the value of the dispersion-function and $\beta_m$ the value of the beta-function at the monitor position. From this equation one can immediately see that the measurement of emittance depends on many parameters. This limits the accuracy to which emittance can be calculated, which is generally with a precision no better than around 10%. A number of instruments are capable of measuring the beam profile quite precisely, but in calculating the emittance one also relies on knowledge of the beam optical parameters at the position of the instrument and these are often fraught with considerable uncertainties.

The remainder of this section is devoted to the measurement of beam size, from which the emittance is then calculated.

### 5.1 Scintillator and optical transition radiation screens

Scintillator screens have been used for nearly a century and are the simplest and most convincing device when one has to thread a beam through a transfer line and into and around an accelerator. The modern version consists of a doped alumina screen that is inserted into the beam and can stand high intensities and large amounts of integrated charge. In its simplest form a graticuled screen is observed using a TV camera. It can deliver a wealth of information to the eye of an experienced observer, but only in a semi-quantitative way. Much can be done about that with modern means of rapid image treatment, but questions concerning the linearity of these screens at high beam densities remain.

Optical transition radiation (OTR) screens are a cheap substitute for scintillator screens. OTR radiation is generated when a charged-particle beam transits the interface of two media with different dielectric constants (e.g. vacuum to metal or vice versa) [21]. Since this is a surface phenomenon, the screens can be made of very thin foils that reduce beam scattering and minimize heat deposition. The radiation produced is emitted in two cones around the angle of reflection for backward (vacuum to metal) OTR so that if the foil is placed at 45° to the beam direction, the radiation produced is at 90° to the beam direction. In addition, two cones of forward OTR (metal to vacuum) are produced around the beam direction (see Fig. 23). The angular distribution of the emitted radiation has a central hole and a peak located at $1/\gamma$. The higher the value of $\gamma$ the sharper the peaks and the more light can be collected, which is why OTR is generally suited to lepton or high-energy hadron machines. However, the experience from modern, Linac-based, fourth-generation light sources (free electron lasers (FELs)) shows that OTR diagnostics fail because of coherence effects in the OTR emission process. As a consequence such machines have reverted to the use of scintillating screen monitors for transverse beam profile measurements, with additional effort to reach high-resolution imaging [22].

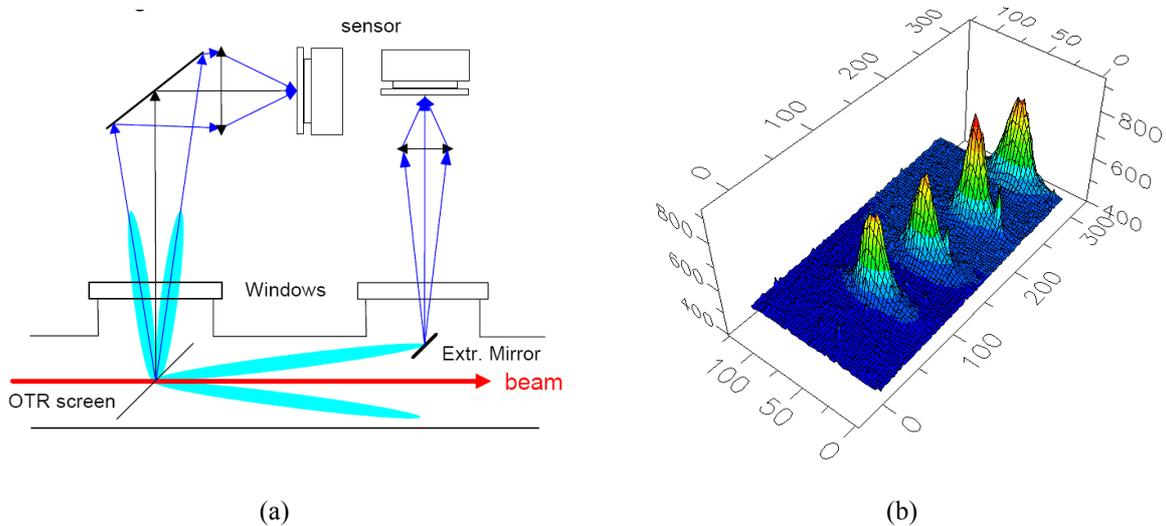

(a)     (b)

**Fig. 23:** (a) Backward and forward OTR patterns with their imaging schemes. (b) An example of 2D OTR images taken every four turns after injection in the CERN-SPS.

### 5.2 SEM grids

Secondary emission (SEM) grids, also known as harps, consist of ribbons or wires that are placed in the beam. As the beam intercepts the grid, secondary emission occurs, leading to a current in each strip that is proportional to the beam intensity at that location. By measuring this current for all strips a beam profile is obtained. SEM grids are the most widely used means of measuring the density profile of beams in transfer lines and low-energy hadron linacs. In addition, sets of three, properly spaced (i.e. with the right phase advance between monitors), allow a determination of the emittance ellipse. What makes them popular is their simple and robust construction, the fact that there is little doubt about the measured distribution, and their high sensitivity, in particular at low energies and for ions. At higher energies they can be considered to be semi-transparent. Amongst their drawbacks are the limited spatial resolution (it is difficult to get the wire spacing much below 0.25 mm) and the rather high cost for the mechanisms and electronics.

## 5.3 Wire scanners

Of all the instruments used for measuring the emittance of circulating beams, wire scanners are considered to be the most trustworthy. They come in two different types; rotative and linear. Rotative wire scanners consist of a thin wire (some tens of micrometres in diameter) mounted on a fork that is attached to a rotating motor (see Fig. 24), while linear scanners use motors that push/pull the wire across the beam. There are two ways of obtaining a beam profile with wire scanners: by measuring the secondary emission current as a function of wire position (similar to the SEM grid acquisition mentioned above) or by measuring the flux of secondary particles created as the beam interacts with the wire. The latter technique is often used for high intensities, where the heating of the wire produces thermal emission that falsifies the secondary emission results. It relies on the use of radiation detectors, typically scintillators followed by photo-multipliers, placed downstream of the wire scanner to detect the $\gamma$-radiation and secondary particles produced when the wire intercepts the beam. To make the flux collected independent of the beam position may require the summation of the signals from two or more detectors positioned around the beam chamber.

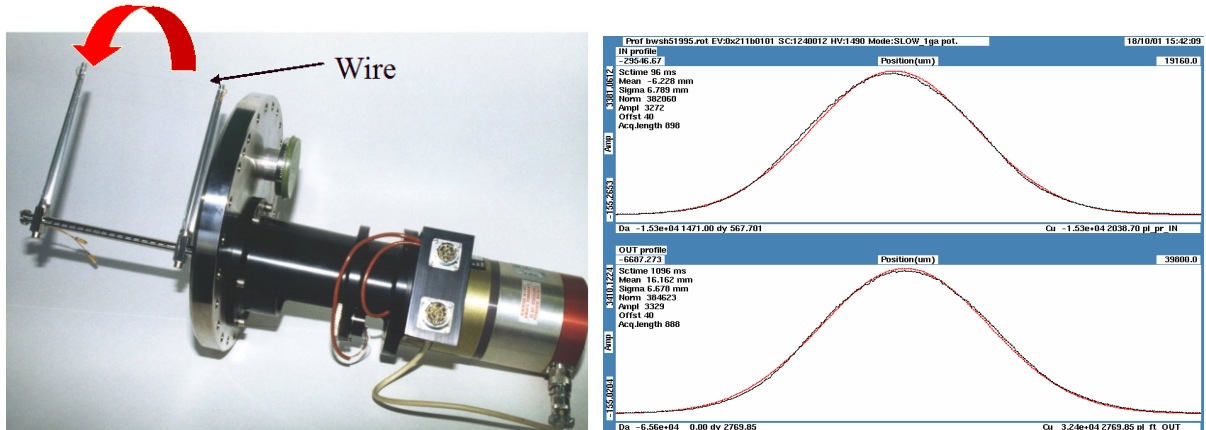

**Fig. 24:** Rotative wire scanner and an example of a wire scanner profile measurement

Fast wire scanners are nearly non-destructive and can be used over a wide range of energies. Their spatial resolution can reach the micrometre range and, with fast gated electronics, can provide the profile of individual bunches. Their great sensitivity also allows them to be used for the study of beam halos.

## 5.4 Residual gas monitors

### 5.4.1 Ionization profile monitors

Ionization profile monitors are used in many high-energy accelerators in order to reconstruct transverse beam distributions (see e.g. Ref. [23]). The signal results from the collection of either the ions or the electrons produced by the beam ionizing the small amount of residual gas in the vacuum chamber. These ions or electrons are accelerated using a bias voltage of several kilovolts and collected on a micro channel plate (MCP). The avalanche of electrons produced by the MCP then either hits a phosphor screen, giving an image of the beam profile that can be monitored using a charge-coupled device (CCD) camera (see Fig. 25), or impinges on a strip detector that can be read-out to give a profile. Due to their rigidity, ions are less sensitive to the distorting effects of the space charge from the circulating beam, but their slow drift time, even with high bias voltages, means that they spend a long time in this beam field, making it difficult to analyse rms beam dimensions smaller than one millimetre. In order to use electrons to produce an image, a transverse magnetic field needs to be added, around which the electrons spiral on their way to the MCP. This eliminates, to a large extent, the space charge effects of the beam and allows sharper images to be produced than with ions. This additional magnetic field, however, is also seen by the beam and has to be compensated by corrector magnets either side of the ionization profile monitor.

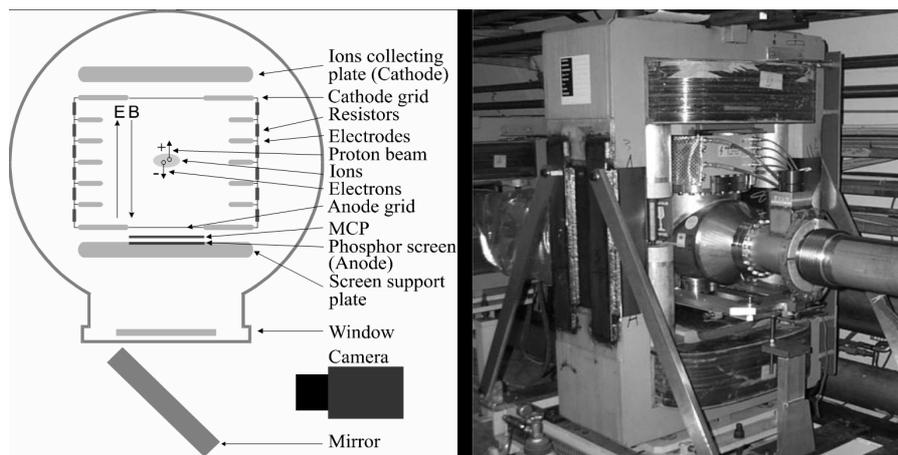

**Fig. 25:** Principle of a residual gas ionization profile monitor and an example from the CERN-SPS

### 5.4.2 Luminescence monitors

Luminescence monitors (see e.g. Ref. [24]) also rely on the interaction of the beam with a gas in the vacuum chamber. The traversing beam excites electrons in the gas molecules from the ground state to a higher energy level. Once the beam has passed the electrons return to the ground state and emit photons of a precise wavelength. In the case of nitrogen the dominant photon wavelength is 391.3 nm, corresponding to light at the lower end of the visible range, for which many detectors are available. In general, the residual gas alone does not produce enough photons for accurate imaging and hence a local pressure bump is usually created by injecting a small amount of gas to enhance the photon production. The principle of luminescence monitoring and a schematic layout of such an instrument are shown in Fig. 26.

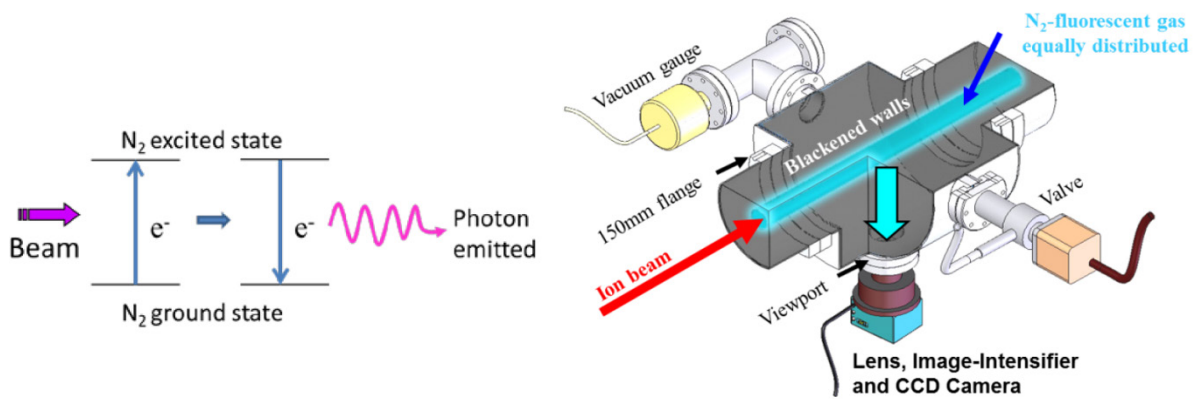

**Fig. 26:** Principle of luminescence monitoring and an example set-up

Most users consider both the residual gas ionization and luminescence profile monitors to be semi-quantitative and not to be relied upon for absolute emittance measurements, even after calibration against some other instrument such as a wire scanner. Their virtual transparency for the beam, however, make them useful for the continuous online tracking of beam size.

### 5.5 Synchrotron radiation monitors

Synchrotron radiation monitors are limited to highly relativistic particles and offer a completely non-destructive and continuous measurement of the two-dimensional density distribution of the beam. These monitors make use of the light produced when highly relativistic particles are deflected by a magnetic field. They are therefore usually positioned to make use of parasitic light produced by a dipole magnet in the machine or behind a purpose-built 'wiggler' or 'undulator' magnet in which the beam is deflected several times to enhance photon emission.

The most common way of measuring the beam size with synchrotron radiation is to directly image the extracted light using traditional optics and a camera. The spatial resolution of such systems is usually limited by diffraction and depth-of-field effects. If the beam is sufficiently relativistic then the photon emission extends into the hard X-ray region of the spectrum and X-ray detectors can be used, for which diffraction effects can be almost disregarded. High-brilliance third-generation synchrotron light sources use special X-ray optics (e.g. pinhole, compound refractive lens, coded aperture, Fresnel zone plate, etc.) to achieve the required resolution ($\approx 1$ μm) for their small beam sizes. A way to overcome the resolution limits in the optical range is to use interferometry or point spread function analysis of the π-polarization of the visible synchrotron light [25, 26].

## 6 Beam loss monitoring

Beam loss monitors (BLMs) have three main uses in particle accelerators.

i) Damage prevention—Beam loss may result in damage to accelerator components or the experimental detectors. One task of any BLM system is to avoid such damage. In some accelerators it is an integral part of the protection system, signalling the beam abort system to fire if a certain loss rate is exceeded. This is of vital importance to the new generation of superconducting accelerators, for which even fairly small beam losses in the superconducting components can lead to magnet quenches.

ii) Diagnostics—Another task of BLM systems is to identify the position (and time) of unacceptable beam losses and to keep the radiation level in the accelerator and its surroundings as low as possible.

iii) Luminosity optimization—BLMs can also help in the tuning of the machine in order to produce the long lifetimes necessary for improved luminosity.

The job of the BLM system is to establish the number of lost particles at a certain position within a specified time interval. Most BLM systems are mounted outside the vacuum chamber, so that the detector normally observes the shower caused by the lost particles interacting in the vacuum chamber walls or in the materials of the magnets. The number of detected particles and the signal from the BLM should be proportional to the number of lost particles. This proportionality depends on the position of the BLM with respect to the beam, the type of lost particles, and the intervening material. It also, however, depends on the momentum of the lost particles, which may vary by a large amount during the acceleration cycle. One has to distinguish between two types of losses:

i) fast losses—where a large amount of beam is lost in a very short time;

ii) slow losses—where partial beam loss occurs over some time (circular machines) or distance (linac, transport lines).

In storage rings, the lifetime is defined by slow losses. There are many reasons for these losses and a BLM system is very helpful for finding out what is happening in the machine. In superconducting accelerators a BLM system can also prevent beam loss-induced quenches caused by these slow losses.

The fact that BLM systems have to cover both of these cases means that they are required to function over a very large dynamic range, typically in the region of $>10^6$.

## 6.1 Long ionization chambers

In 1963, Panowsky [27] proposed a BLM system for SLAC that consisted of one long (3.5 km) hollow coaxial cable filled with Ar (95%) + $CO_2$ (5%), mounted on the ceiling along the LINAC, about 2 m from the beam (PLIC). When a beam loss occurs, an electrical signal is produced that propagates to both ends of the cable. Position sensitivity is achieved by comparing the time delay between the direct pulse from one end and the reflected pulse from the other. The time resolution is about 30 ns (~8 m), which, for shorter versions, can be reduced to about 5 ns. This principle of space resolution works for linear accelerators and transport lines with a bunch train much shorter than the machine and with relativistic particles. For particles travelling significantly slower than the signal in the cable the resolution of multiple hits in the cable becomes difficult. In this case, and for circular machines, it is necessary to split the cable. Each segment has to be read out separately, with a spatial resolution that becomes approximately equal to their length.

## 6.2 Short ionization chambers

Short ionization chambers are used in many accelerators (see e.g. Ref. [28]). They are more or less equally spaced along the accelerator with additional units at special positions such as aperture restrictions, targets, collimators, etc. The chamber provides some medium with which the secondary particles created by the beam loss can interact, typically a gas such as nitrogen or argon. This interaction produces electron–ion pairs that are collected by a series of high-voltage gaps along the length of the chamber. The resulting current is then measured and is proportional to the beam loss at the location of the monitor. An example of a CERN-LHC ionization chamber is shown in Fig. 27.

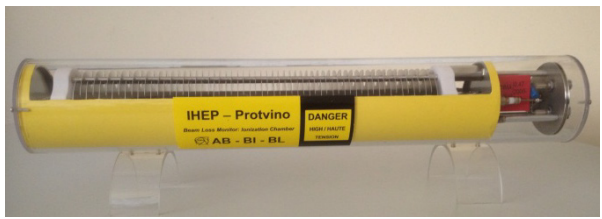
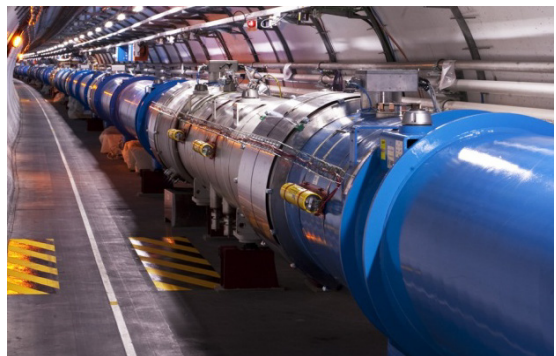

**Fig. 27:** A CERN-LHC ionization chamber used for beam loss monitoring

## 6.3 Scintillation counters

In the case where losses occur in a machine without a full BLM system, a plastic scintillator with photomultiplier readout is often temporarily installed. Such systems have well-known behaviour, but radiation damage to the plastic scintillator restricts their long-term use. Liquid scintillators are not susceptible to such damage and have been installed in some accelerators [29, 30]. Such BLMs can be very fast, with pulse rise times of around 10 ns, but suffer from drift in the photomultiplier gain.

### 6.4 Aluminum cathode electron multipliers

In such detectors the sensitivity of photomultipliers to ionizing radiation is increased by replacing the photocathode with an aluminium foil. This foil then works as a secondary electron emitter when irradiated. A BLM system consisting of aluminum cathode electron multipliers (ACEMs) is installed in the CERN-PS and PS-Booster [31]. It is very fast, with signal rise times in the order of 10 ns, but is rather expensive since the ACEM is not a standard tube supplied by photomultiplier manufacturers.

### 6.5 PIN photodiodes

For circular electron accelerators, which emit hard synchrotron radiation, it is difficult to distinguish between the beam loss distributions and the synchrotron radiation background using traditional BLM techniques. In DESY-HERA, an electron–proton collider, the warm electron ring, and a superconducting proton ring were in the same tunnel. Protection of the superconducting proton beam magnets from beam loss-induced quenches therefore relied on a BLM system that saw only the proton beam losses and not the synchrotron radiation background. In this case back-to-back PIN photodiodes were used to distinguish between the hadronic shower created by beam losses and the synchrotron radiation [32]. The charged particles interacted with both photodiodes, giving a coincidence signal, while the photons were absorbed by the first diode. In contrast to the charge detection of most other BLM systems, PIN photodiode detection depends on counting coincidences, with the count rate proportional to the loss rate so long as the number of overlapping coincidences is small.

### 6.6 Optical fibres

Cherenkov light created in long optical fibres can be used in the same way as a single long ionization chamber to reduce the costs of a BLM system while maintaining full coverage. It allows real-time monitoring of loss location and loss intensity as for PLICs (Section 6.1).

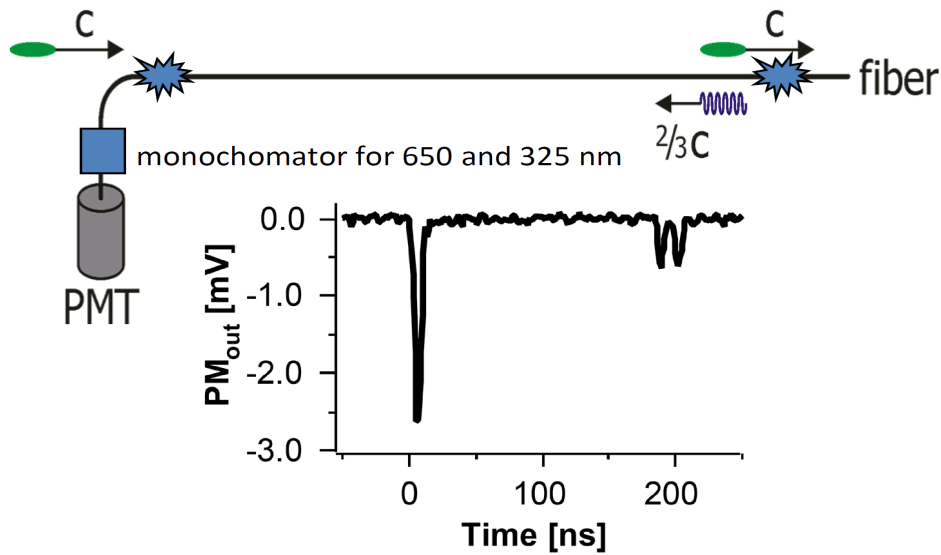

**Fig. 28:** A time measurement provides the position of the beam loss along the fibre while the amplitude gives the amount lost. Due to dispersion, the emission at 650 nm reaches the photodetector earlier than the emission at 325 nm, thus producing the double peak (Tesla-Report 2000-27).

A Cherenkov detector is nearly insensitive to X-rays because the Cherenkov effect only occurs when the velocity of a charged particle traversing a dielectric medium is faster than the speed of light in that medium. The fast response of the Cherenkov signal is detected with photomultipliers at the end of the irradiated fibres (Fig. 28). A time measurement provides the position of the loss along the fibre while the integrated light is proportional to the size of the loss. A longitudinal position resolution of

20 cm (= 1 ns at $v = 0.66c$) is possible with this technique. High-purity quartz fibres (Suprasil®) are often used for such installations as they only produce Cherenkov emission (no scintillation) and are radiation hard, capable of withstanding tens of MGy of irradiation. Scintillating fibres, on the other hand, give about 1000 times more light output but are much more sensitive to radiation damage.

## 7  Longitudinal profile diagnostics

The time structure of a charged particle beam is related to the characteristics of the accelerating field. Electrostatic accelerators produce DC particle beams that have no variation in time. RF accelerators generate beams with a time structure defined by the frequency of the electromagnetic field in use. Currently RF accelerators may operate at frequencies up to tens of gigahertz, which means that particle bunches can naturally be as short as a few picoseconds or less. For some applications, such as FELs or TeV linear colliders, the bunch length is further compressed down to 10–100 fs. Moreover, since the late 1990s, the use of ultra-short intense laser pulses has contributed to the development of innovative acceleration schemes such as laser plasma wakefield acceleration (LPWA). Here, the accelerating field is generated by the interaction of a laser beam with a gas cell. The resulting plasma wave has a very high electric field that may oscillate at frequencies upto terahertz with the corresponding bunch length in the order of a few femtoseconds.

In order to measure the longitudinal structure of particle beams, many different instruments have been developed during the last three decades. They can be regrouped into four different categories:

i) direct beam observation, where the longitudinal structure of the beam is measured by means of fast detectors;

ii) detection of coherent radiation, where the bunch length can be retrieved by measuring the radiated power spectrum for wavelengths of the order or longer than the bunch length;

iii) RF manipulation, where the longitudinal beam profile is encoded into the modulation of its transverse spatial profile;

iv) sampling techniques, where a short laser pulse is used to scan through the longitudinal bunch profile.

### 7.1  Direct beam observation

For relatively long bunches, the beam time structure can be measured using wall current monitors (WCMs) [33] and fast-sampling oscilloscopes. WCMs have already demonstrated bandwidths in excess of 10 GHz and state of the art oscilloscopes are improving their performance every year with sampling rates as high as 80 GSa/s now available. To provide even better time resolution optical methods need to be used. An optical replica of the longitudinal beam distribution must first be generated through the use of synchrotron radiation, transition radiation, diffraction radiation, or Cherenkov radiation, with the corresponding light pulses measured by an appropriate detector.

#### 7.1.1  *Single photon counting*

A way of providing both time resolution in the order of tens of picoseconds and a high dynamic range is through the use of time-correlated single-photon counting [34]. This principle is illustrated schematically in Fig. 29. The system is well adapted to ring accelerators, where each particle bunch passes the detector once every revolution with the light attenuated such that on average less than one photon reaches the detector per bunch passage. The detector typically used is an avalanche photo diode (APD) that produces an electrical pulse when it detects an incoming photon. This is time-stamped by a time-to-digital converter (TDC) and a histogram of arrival times is created. In order to construct a meaningful bunch profile the data have to be collected over thousands of turns. The longer the acquisition, the more counts are added to the histogram and the higher the dynamic range of the measurement.

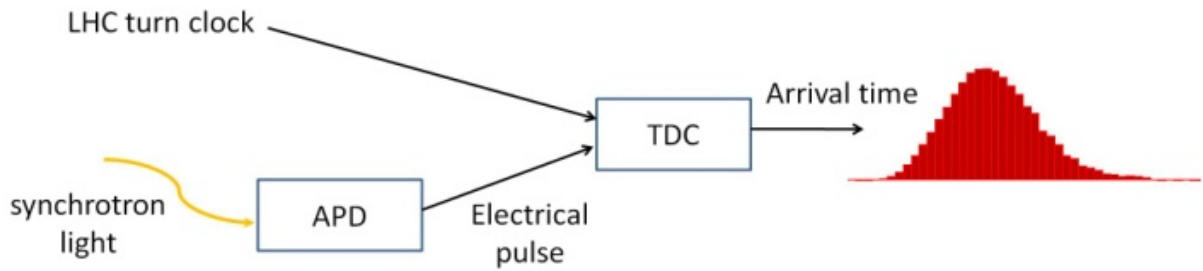

Fig. 29: Schematic of a single-photon-counting longitudinal density monitor

An example of a typical profile measured on the LHC at CERN is shown in Fig. 30, illustrating the large dynamic range that can be achieved when applying all the appropriate signal corrections. A long integration time of several minutes is necessary in order to make a high dynamic range profile showing small satellites at the $10^{-4}$ level of the main bunches.

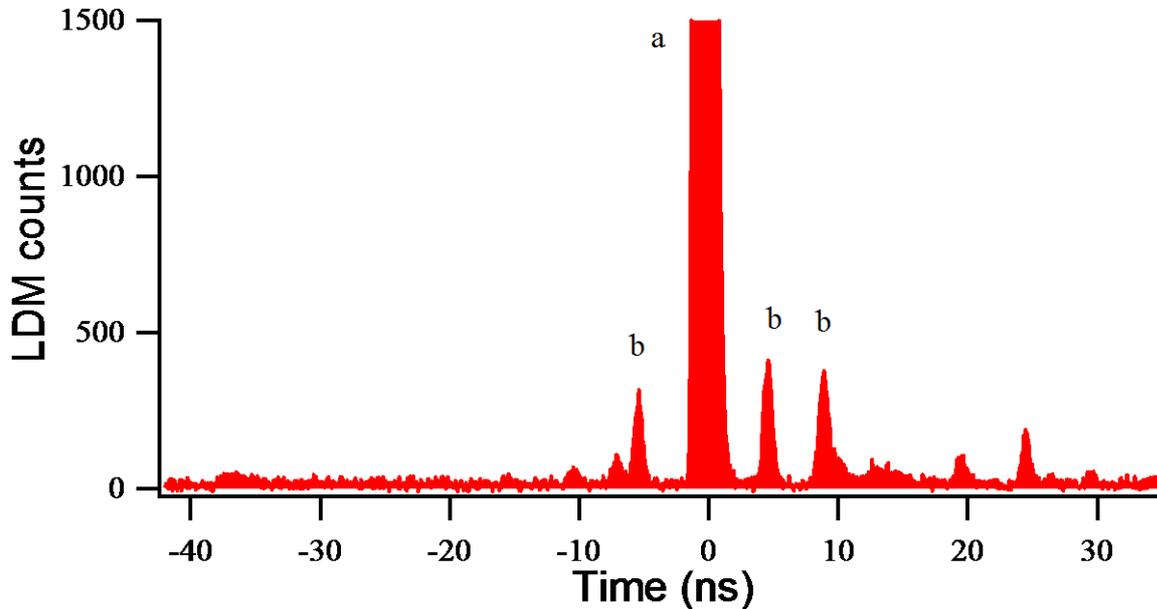

Fig. 30: Longitudinal density measurement in the LHC showing (a) a main proton bunch with peak at $1.3 \times 10^5$ counts and (b) small satellite bunches.

### 7.1.2   Streak camera

Another very popular method to measure longitudinal bunch profiles using optical signals relies on the use of streak cameras [35]. In order to achieve a good time resolution, the photons from the radiation to be analysed are converted to electrons, which are then accelerated and deflected using a ramped, time-synchronized, high voltage (HV) electric field, as shown in Fig. 31. The signal from the electrons is subsequently amplified with an MCP, converted to photons via a phosphor screen, and finally detected using an imager such as a CCD array. In this way the time variation of the intensity of the incoming photon pulse, which is a replica of the longitudinal bunch profile, is converted into a spatial intensity variation. The time resolution of a streak camera is limited by the transverse size of the collimation slit, the initial velocity spread of the photo-electrons, and the dispersion of the optical system. State-of-the-art streak cameras can provide a time resolution better than 500 fs.

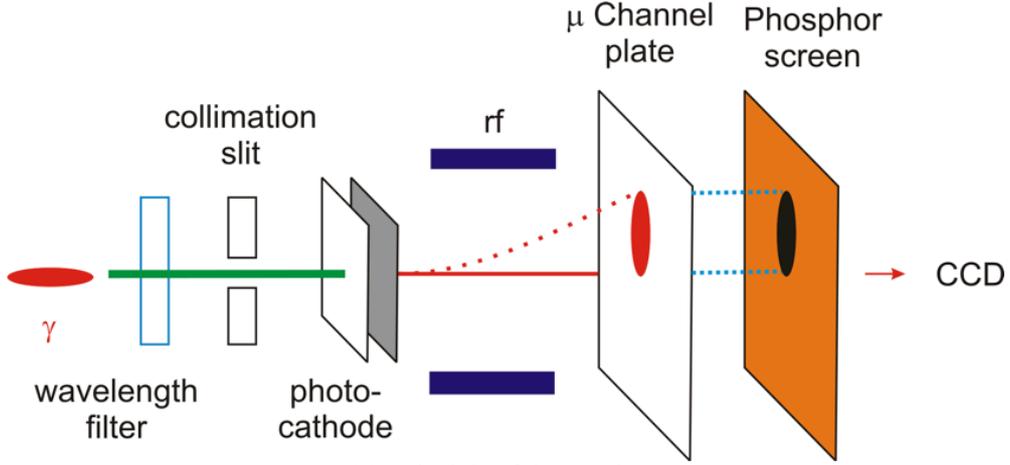

**Fig. 31:** Principle of the streak camera

## 7.2 Coherent radiation

For wavelengths shorter than the bunch length, the particles within the bunch radiate incoherently, with the power emitted proportional to the number of particles. However, for wavelengths equal to or longer than the bunch length, the particles emit radiation in a coherent way with the emitted power dependent on the bunch length and scaling as the square of the number of particles. This can be described by the following equations:

$$\text{incoherent term} \quad \text{coherent term}$$
$$\Downarrow \qquad\qquad \Downarrow$$
$$S(\omega) = S_\text{p}(\omega)\left[N + N(N-1)F(\omega)\right] \qquad (10)$$

$$F(\omega) = \left|\int_{-\infty}^{\infty} \rho(s)\, e^{-i\frac{\omega}{c}s}\, ds\right|^2 \qquad (11)$$

where $S(\omega)$ represents the radiation spectrum, $S_\text{p}(\omega)$ the single particle spectrum, $N$ the number of particles and $F(\omega)$ the longitudinal bunch form factor, which depends on the longitudinal particle distribution $\rho(s)$.

Measuring the power spectrum therefore allows the form factor to be calculated from which an indirect estimate of the bunch length is possible. This method is relatively simple to implement and has already demonstrated its capacity for the measurement of extremely short bunches.

The advantage of coherent radiation monitors is that their resolution does not have any theoretical limit. What is obtained, however, is the form factor rather than the longitudinal distribution. In order to reconstruct the longitudinal bunch profile it is necessary to perform an inverse-Fourier transformation using phase recovery algorithms making use of the Kramers–Kronig relation.

The detection can be based on an autocorrelation technique using either Michelson or Martin–Puplett interferometers. Scanning the interferometer over the frequency range of interest allows a measurement of the radiated power spectrum, $S(\omega)$, from which the longitudinal bunch profile can be reconstructed. An example of such a measurement performed on the Flash FEL facility at DESY [36] is depicted in Fig. 32 showing the measured power spectrum and the reconstructed bunch profile compared with a streak camera measurement. Single-shot measurement devices have also been developed using gratings combined with a large array of detectors [37].

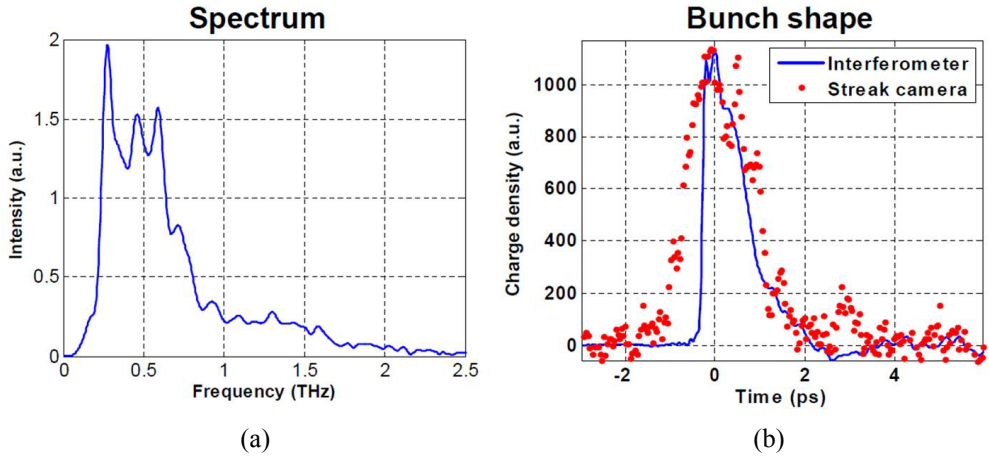

| (a) | (b) |

**Fig. 32:** (a) Measured coherent synchrotron radiation spectrum; (b) reconstructed bunch shape

### 7.3 RF manipulations

The principle of RF manipulation techniques is to encode the longitudinal structure of the beam into spatial information, which is easier to measure. Two such techniques, a bunch shape monitor for hadron linacs and an RF deflecting cavity for FEL or e$^+$–e$^-$ linear colliders, are described below.

#### 7.3.1 Bunch shape monitor

For non-relativistic beams ($\beta \ll 1$) the electromagnetic field of a bunch is not purely transversal and therefore does not represent its longitudinal charge distribution. Hence, a bunch shape monitor based on secondary emission has been developed to measure the longitudinal bunch profile in a non-invasive way [38]. These monitors are used in many proton, H$^-$, and ion linacs. Their principle of operation is presented in Fig. 33.

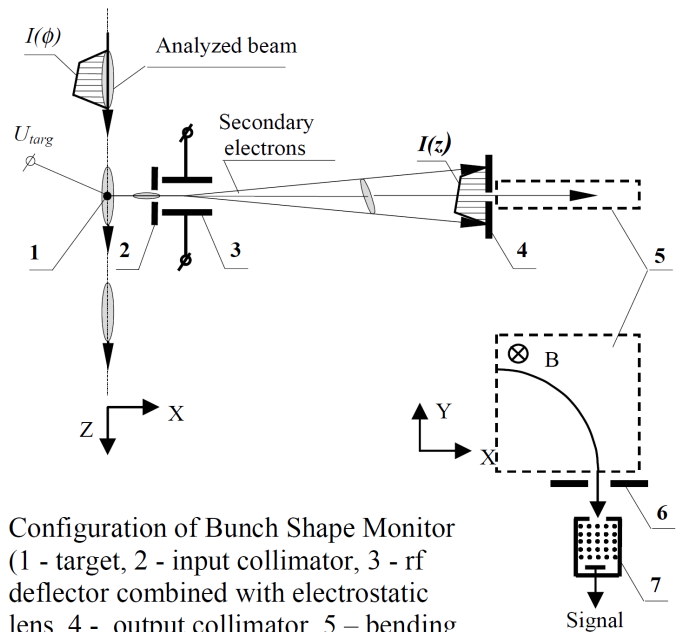

Configuration of Bunch Shape Monitor (1 - target, 2 - input collimator, 3 - rf deflector combined with electrostatic lens, 4 - output collimator, 5 – bending magnet, 6 – collimator, 7 – Secondary Electron Multiplier).

**Fig. 33:** Principle of operation of the bunch shape monitor

As the beam hits a thin metal wire, low-energy secondary electrons of a few eV are released almost instantaneously. These electrons are quickly accelerated away by applying a negative bias voltage ($U_{\text{targ}} \approx -10$ kV) on the wire. The secondary electron beam is then collimated and deflected by a time-varying transverse RF field. Similarly to the principles behind the streak camera, the arrival time of the electrons is thereby transformed into a spatial distribution that can be measured by various detectors, such as a phosphor screen coupled to a CCD camera, a multichannel electron detector, or a scanning slit and collector. State-of-the-art BSMs have achieved resolutions in the order of some 10 ps.

In the special case of an H⁻ beam, the detached electrons originating from the dissociation of the H⁻ ions on the target wire contribute to the background signal. Energy separation by an additional spectrometer behind the second slit can be used to reduce this background, enabling measurements of the longitudinal halo with a dynamic range of $10^5$.

The energy deposition of the beam in the wire must remain small in order to avoid the creation of thermal electrons and ultimately the melting of the wire. This clearly imposes operational limitations when using a BSM with high-charge beams.

### 7.3.2 RF deflecting cavities

The principle of operation of a longitudinal bunch profile monitor using an RF deflecting cavity [39] is shown in Fig. 34.

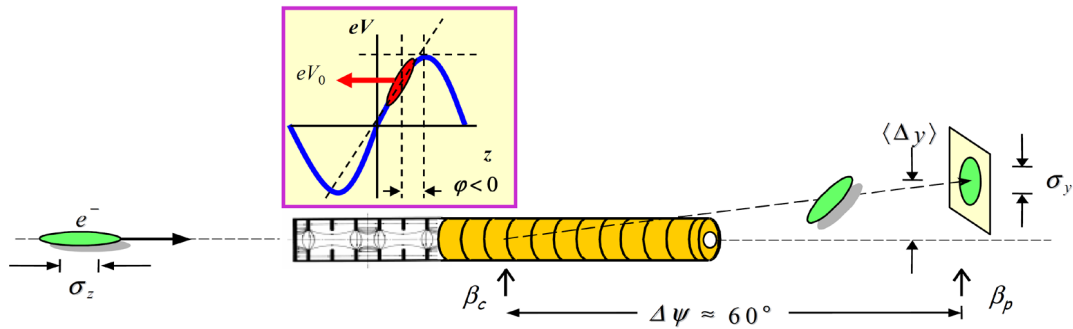

**Fig. 34:** Principle of operation of a longitudinal bunch profile monitor using an RF deflector

Again, the idea is to transform a longitudinal bunch distribution into a transverse distribution. The cavity is designed to provide a time-varying deflecting field. By adjusting the arrival time of the bunch with respect to the phase of the electric field, the particles will experience different transverse kick strengths depending on where they are located along the bunch. By measuring the transverse beam size downstream, one can then reconstruct the longitudinal bunch profile. Assuming a deflection in the vertical plane, the vertical beam size $\sigma_y$ can be expressed as

$$\sigma_y = \sqrt{\sigma_{y_0}^2 + \sigma_z^2 \beta_c \beta_p \left( \frac{2\pi}{\lambda} \frac{eV_0}{E_0} \sin(\Delta\Psi) \cos(\varphi) \right)^2} \quad (12)$$

where $\sigma_{y0}$ is the vertical beam size measured on the screen when the deflector is turned off, $\sigma_z$ the bunch length, $\beta_c$ and $\beta_p$ the beta-function at the position of the cavity and the screen respectively, $\Delta\psi$ the betatron phase advance between the cavity and the screen, $\lambda$ the RF cavity wavelength, $V_0$ the RF voltage in the cavity, $\varphi$ the phase of the RF field and $E_0$ the beam energy.

RF deflectors have demonstrated the capability of measuring very short bunches down to a few femtoseconds but have the drawback of being destructive to the measured beam and having a very high cost.

## 7.4 Sampling techniques

Sampling techniques rely on the use of very short laser pulses. Over the past 15 years, several monitors have been tested using different processes such as electro-optic Pockel and Kerr effects [40] or Compton scattering [41]. Whatever the physics process involved, these techniques always require laser pulses shorter than the particle bunch length and a precise laser-to-beam synchronization. Only the electro-optic techniques will be described in more detail here.

### *7.4.1 Principle of electro-optic diagnostics*

The aim of electro-optic (EO) longitudinal diagnostics is to accurately measure the temporal profile of the Coulomb field of a relativistic particle beam using the optical non-linearity induced in an electro-optic crystal. This only provides an accurate replica of the longitudinal bunch profile for highly relativistic beams, where the accompanying Coulomb field has an opening angle much smaller than the bunch length. The coulomb field is probed by placing an electro-optic crystal adjacent to the particle beam (Fig. 35), with no need for the beam to traverse the crystal, making this a completely non-intercepting technique.

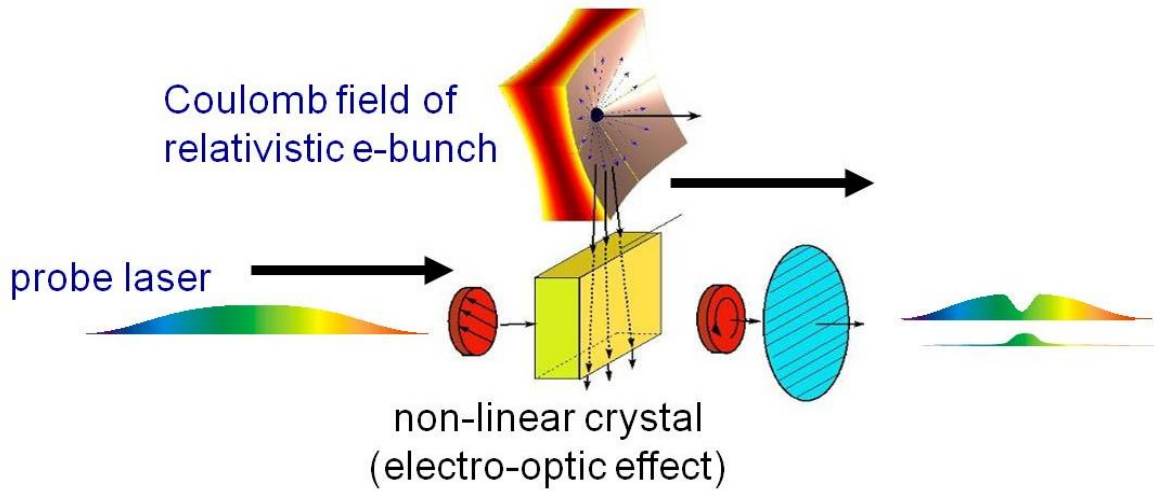

**Fig. 35:** Principle of electro-optic detection

The Coulomb field sweeping through the crystal renders the material bi-refringent during its transit. This birefringence is probed using a chirped (or sometimes ultra short) optical probe laser pulse that is passed through the crystal parallel to the particle beam axis in synchronism with the particle bunch. The birefringence results in a rotation of the polarization of the optical pulse. This can be sensitively detected using optical techniques to yield a temporal evolution of the Coulomb field, hence measuring the longitudinal charge density profile of the bunch.

### *7.4.2 Encoding and decoding techniques*

The temporal resolution of the longitudinal profile obtained with EO detection depends both on the processes occurring within the non-linear crystal during the encoding, and by the method of decoding the temporal information from the resulting optical pulse. For all techniques the time resolution is also limited by the Lorentz factor of the particles, which dictates how faithful a reproduction of the bunch charge density profile the Coulomb field actually is. In practice, for the GeV electron beams for which such techniques are typically used, this is not a limitation.

The encoding limitations arise from the EO crystal response, and are best viewed as a frequency cut-off, rather than directly as a temporal resolution. For example, a 0.2 mm-thick ZnTe crystal can efficiently detect frequencies up to 2.8 THz, while a similar thickness GaP crystal works up to about

8 THz. For ultra-short bunches with a frequency above this cut-off, the resulting profile will no longer be an accurate reproduction of the traversing Coulomb field.

For decoding the information, the two techniques most widely used are spectral decoding and single-shot temporal decoding (Fig. 36). In both cases the optical pulse from an ultrafast laser is synchronized to the particle bunch train. Such an ultra-short pulse, by definition, contains a certain spread in wavelengths. This fact can be used to stretch the optical pulse, through the use of gratings or dispersive media, so that it is longer than the bunch length to be measured, and has a time to wavelength correlation. The stretched laser pulse passes through a polarizer and is focused onto the electro-optic crystal. The birefringence induced in the EO crystal during the passage of the bunch is translated into a change in polarization and subsequently, through the use of polarizers, into intensity modulation on the laser pulse.

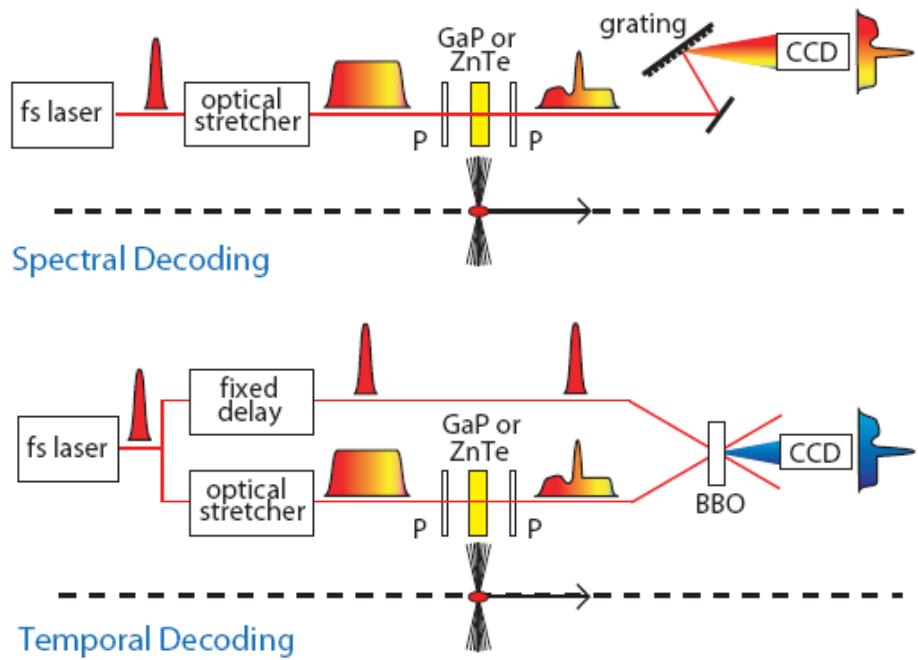

**Fig 36:** General layout for spectral decoding (top) and temporal decoding (bottom)

Spectral decoding makes use of the fact that there is a correlation between time and wavelength to resolve the bunch profile using a spectrometer, and can measure bunch lengths down to 500 fs.

For temporal decoding the laser beam is split into two, providing a probe and a gate pulse. The probe, as for spectral decoding, is stretched, polarized and focused onto the crystal. The resulting intensity modulated probe is then passed through a β-Barium Borate (BBO) crystal at the same time as the short gate pulse. By selecting the right angles of incidence the short gate pulse can be made to scan across the longer probe pulse within the BBO crystal, producing frequency-doubled light with the same intensity profile as the probe.

Temporal decoding has significantly better time resolution capabilities than spectral decoding, and has been used to measure electron bunches shorter than 100 fs at DESY-FLASH (Fig. 37) [42]. However, the price to pay is the requirement for much higher laser power.

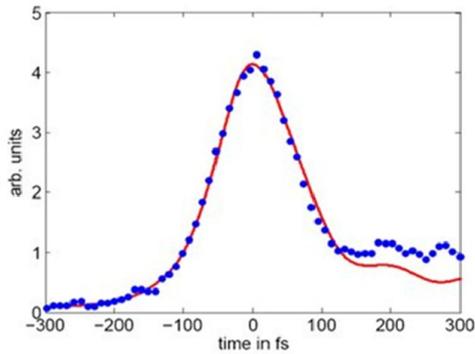 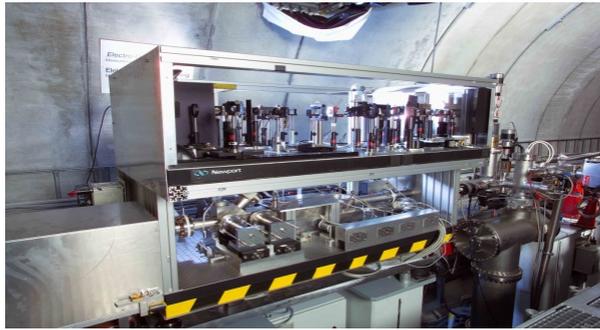

**Fig. 37:** Shortest measured electron bunch profile using temporal decoding at DESY FLASH. The optimum fitted Gaussian curve sigma is 79.3 ± 7.5 fs.

# 8 Some examples of beam diagnostics

This section is meant to serve as general entertainment for those readers who have made it this far! Two examples from CERN-LEP operation have been selected, and show how difficult it can be to interpret primary measurements and decide on the right actions for solving a problem in an accelerator.

## 8.1 The CERN-LEP beam does not circulate!

The LEP accelerator had a very regular operation schedule. Every year it was used for about eight months for physics, followed by a four-month maintenance and upgrade shutdown. During this shutdown major intervention work was sometimes carried out on the machine. At the next start-up it was often expected that typical problems, such as inverted magnet polarities, would have to be overcome. One year the start-up was particularly bad, with neither the electron beam nor the positron beam capable of being made to circulate. Several hours were devoted to checking all vacuum conditions, power supply currents, settings of the radio frequency system, injection deflectors, and so on, but nothing indicated a severe problem. Finally, people started to look in detail at the measured beam trajectory from the injection point onwards. A typical example for the positron beam is shown in Fig. 38.

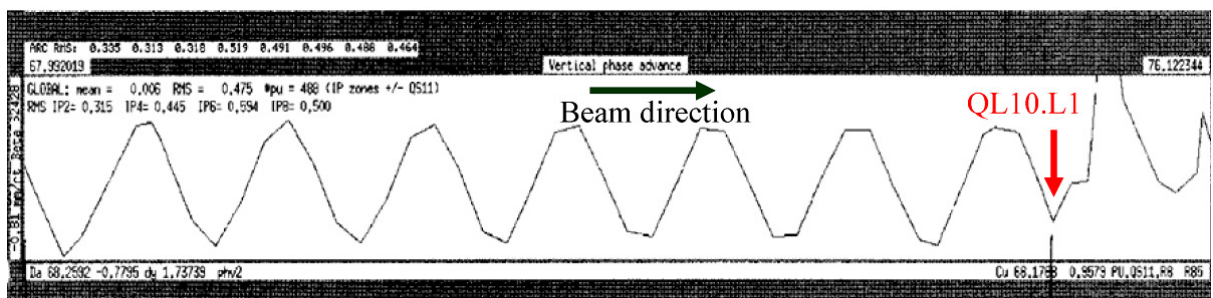

**Fig. 38:** Measurement of the LEP phase advance when beams did not circulate

What is actually shown in Fig. 38 is the phase advance from one beam position monitor to the next, as calculated from the measured beam trajectory. At a particular quadrupole (QL10.L1) the regular pattern is distorted. Additional measurements also indicated that most of the beam was lost at this point. The first conclusion was to suspect a problem with this quadrupole. People went in, measured the current in the quadrupole, checked its polarity, inspected its coils, but could not find anything abnormal.

The indications of the beam measurements, however, clearly pointed to a problem at this location. After many discussions and potential hypotheses it was decided to open the vacuum chamber. It should be noted that this was a major intervention, causing a stop of the accelerator for at least one day. One can understand the surprise of the intervention team when they looked into the open vacuum chamber and saw a beer bottle!!! During the shutdown intervention, somebody had sabotaged the LEP accelerator and inserted a beer bottle into the beam pipe (Fig. 39)! What had upset the operation team most at the time was the fact that it was a very unsociable form of sabotage—the bottle was empty!

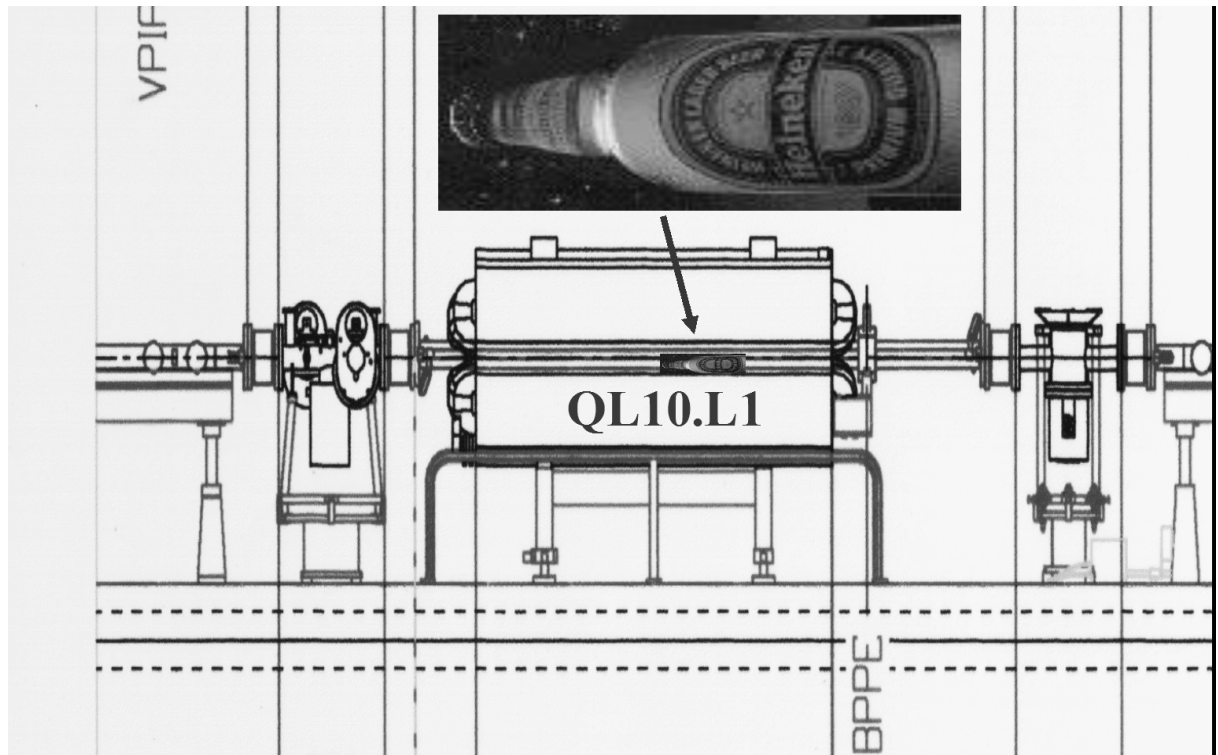

**Fig. 39:** The mystery of the beam circulation problem in LEP is solved!

## 8.2 The beam gets lost during the beta squeeze

This is another of the stories from LEP operation that took several hours using beam diagnostics to solve. The problem in itself is pretty complex, and therefore requires some additional explanations beforehand.

The acceleration of the particle beams and the change of the lattice function in the insertion regions in order to get smaller values of the beta-function at the crossing point (hence higher luminosity) are so-called 'dynamic processes'. The presence of the beam requires that all actions are well synchronized. For example, the power converters of all relevant magnetic circuits have to be controlled such that beam parameters such as the closed orbit, tunes, and chromaticities stay within tolerance during the dynamic process. In order to achieve this, the behaviour of these beam parameters is periodically measured as a function of time and the corresponding power converter tables are updated.

During one period of LEP operation it was found that the beams were lost during the beta squeeze. Shortly before the total loss of the beams a significant beam loss was measured. As standard practice when encountering such problems, the Engineer In Charge (EIC) launched a new machine cycle with diagnostic facilities such as 'tune history' (the measurement of the betatron tunes as a function of time—see Section 4) switched on. This indicated that the vertical tune moved out of tolerance during the beta squeeze. Figure 40 shows an excerpt from the actual LEP logbook entry of this event.

> Straight through to 98 GeV.
> At ~ 97–98 GeV e⁻ large vertical oscillation
> OPAL trigger. Maybe a bit too ambitious
> Tunehistory   01-12-40   fill 7065
> → nothing particularly nasty.
> Big radiation spikes in all expts.
>
> 01:40   22 GeV   4QS0   Breakpoint at 93 GeV.
> 640 µA   .234 / .164   5.27 mA
> 93 GeV   4QS0         01-58-36   V rms ~ c
> Tunehistory   01-50-25   fill 7066

Fig. 40: Excerpt from the LEP logbook when beams were lost during the beta squeeze

As a result of this observation, the EIC launched another cycle, but inserted a breakpoint (to stop the accelerator cycle) just before the critical moment in the beta squeeze when the deviation in tune occurred. Having reached that breakpoint the tunes were measured statically and found to be perfectly within tolerance. The beta squeeze was then executed step by step, and to the big surprise of the operations crew, the tunes were found to be correct at all times. The beam had passed the beta squeeze like on an ordinary day! But on the next attempt, without a break in the cycle, the beam was again lost at the same moment, and several people scratched their heads to find an explanation.

Finally, the following measurement was made. The machine was prepared and a breakpoint again inserted just before the critical beam loss. Once this point was reached, the EIC requested the execution of one further step in the beta squeeze. The facility by which one could execute a single step in a dynamic process had the additional feature that one could specify the rate of current change of any machine element. This current rate limitation was changed from 5 A/s (nominal) down to 2.5 A/s on consecutive steps. The corresponding tune history (the result from the vertical plane is plotted on the lower graph) is shown in Fig. 41.

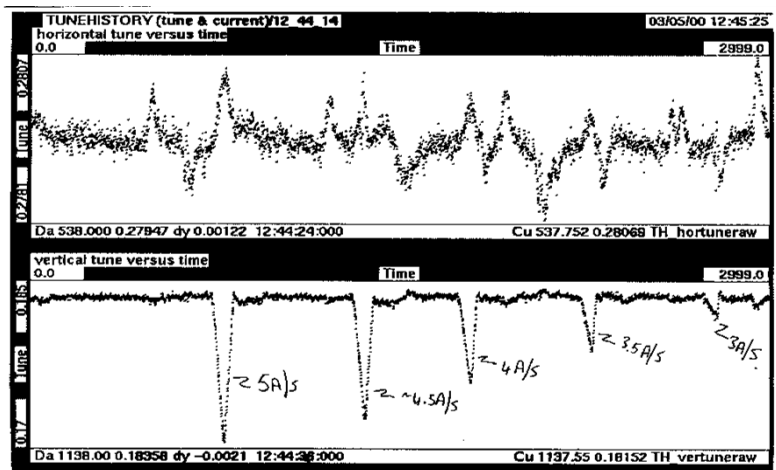

Fig. 41: The LEP tune history during the beta squeeze for various power converter ramp rates

One can clearly see that a huge (negative) tune excursion occurred when the step was executed at the nominal rate. This observation led the EIC to the right conclusion, which was that one of the power supplies was able to deliver the demanded current statically, but not dynamically. When this was discussed with experts from the power converter group, they indicated that the power supplies for the superconducting insertion quadrupoles were built as two blocks in series, each of them able to deliver the necessary current (each block typically 1000 A/10 V). Both of these blocks were required to have enough voltage margins to enforce a current change against the inductance of the quadrupole coil. This then explained the whole story. One of these blocks was faulty, but since the remaining working block could deliver its static current, it was not detected by an alarm or surveillance circuit. If the dynamic rate was too high, however, this single block could not provide enough current leading it to lose synchronism with the other power converters. This resulted in the large tune change observed and ultimately the total beam loss.

These two examples show the enormous potential of beam instrumentation if they are used in the right combination by intelligent people.